\documentclass[%
11pt,
preprint,
 nofootinbib,
 notitlepage,
 amsmath,amssymb,
 aps,
prd,
longbibliography
]{revtex4-2}

\usepackage{graphicx}
\usepackage{dcolumn}
\usepackage{bm}
\usepackage[bookmarks=true,colorlinks=true,linkcolor=blue,unicode=true]{hyperref}
\usepackage[mathlines]{lineno}
\usepackage{xcolor} 
\usepackage{color} 

\newcounter{YJC}

\begin{document}


\title{Wilson loops with oppositely oriented plaquettes as a probe of center vortex structure}

\author{Ji-Chong Yang}
\email{yangjichong@lnnu.edu.cn}
\thanks{corresponding author}
\author{Xiang-Ning Li}
\email{2788249040@qq.com}
\author{Zhan Zhao}
\email{zhaozhan061300@163.com}

\affiliation{Department of Physics, Liaoning Normal University, Dalian 116029, China}
\affiliation{Center for Theoretical and Experimental High Energy Physics, Liaoning Normal University, Dalian 116029, China}

\date{\today}

\begin{abstract}
We study Wilson loops with a nontrivial orientation structure in lattice gauge theory as a probe of the center vortex picture. 
The observable is a single Wilson loop containing two plaquettes with opposite orientations, realized in two geometries referred to as the vertical and parallel configurations. 
The vertical case behaves consistently with expectations from the vortex picture. In contrast, the parallel configuration shows a deviation from the naive area-law expectation which cannot be explained solely by the opposite orientations of the plaquettes. 
We introduce a simple qualitative vortex model which accounts for this behavior and shows that the observed effect can still be understood within the vortex framework.
\end{abstract}

\maketitle


\section{\label{sec1}Introduction}

Color confinement remains one of the central unresolved problems in quantum chromodynamics~(QCD). 
In lattice gauge theory, confinement is characterized by the area-law behavior of large Wilson loops and the emergence of a linearly rising static quark-antiquark potential at large separations. 
Among the proposed microscopic mechanisms for confinement, the center vortex picture has emerged as one of the most compelling descriptions. 
In this picture, the relevant long-distance degrees of freedom of the Yang-Mills vacuum are extended topological objects known as ``center vortices'' which carry quantized magnetic flux associated with the center group $\mathbb{Z}_N$ of the gauge group $SU(N)$ after the gauge field is projected on to the center group~\cite{tHooft:1977nqb,Mack:1978rq,Cornwall:1979hz}. 
A Wilson loop linked with a vortex acquires a nontrivial center phase, and a random distribution of such vortices leads naturally to an area law, providing a direct geometric mechanism for confinement~\cite{Greensite:2003bk}. 

Lattice QCD has strong quantitative support for the center vortex picture. 
Using gauge-fixing procedures such as maximal center gauge followed by center projection, vortex degrees of freedom can be identified and isolated~\cite{DelDebbio:1996lih,Langfeld:1997jx}. 
A series of numerical studies has demonstrated that the string tension is almost entirely reproduced by vortex-only configurations, while vortex-removed configurations lose confinement altogether~\cite{DelDebbio:1998luz,Faber:1997rp}. 
Beyond confinement, vortices have been shown to play a crucial role in chiral symmetry breaking~\cite{Gattnar:2004gx,Bowman:2010zr,Trewartha:2015nna}. 
Geometric studies further reveal that vortices form extended, percolating structures in the confining phase, supporting a picture of vortex condensation in the QCD vacuum~\cite{Engelhardt:2002qub,Langfeld:2003ev}. 
Besides, it has been shown that the vortex branching also behaves differently between the confined phase and the deconfined phase~\cite{Spengler:2018dxt}.
These results collectively suggest that center vortices are not merely a confinement mechanism but may constitute a unifying infrared structure underlying multiple emergent properties of QCD~\cite{Greensite:2011zz}.

Despite this success, significant challenges remain in the center vortex picture. 
One of the frequently mentioned issues is the dependence of the vortex identification on the gauge fixing.
The true physical topological structure of a field must be gauge-invariant. 
Moreover, due to the non-uniqueness of gauge fixing, procedures such as maximal center gauge suffer from Gribov copy ambiguities, which can affect vortex localization and compromise the robustness of vortex detection~\cite{Bornyakov:2000ig}. 
In addition to research on the effects of Gribov copies and studies on measurement techniques for center vortices, a question worth asking is whether center vortices are real, their existence should not depend on gauge transformations. 
For instance, we know there is a correspondence between center vortices and near-zero Dirac eigenmodes~\cite{Hollwieser:2008tq,Baranka:2024cuf}. 
Moreover, research is being conducted to observe the center-vortex phenomena using gauge-invariant measurement methods~\cite{Greensite:2014gra}, such as persistent homology~\cite{Sale:2022qfn}.
Based on this perspective, the present work considers a gauge-invariant observable, namely the results of Wilson loops with different orientations. 
We find that the expectation value of a Wilson loop does not simply follow an area law, and its behavior for different orientations is consistent with the vortex picture.

The remainder of the paper is organized as follows. 
In Sec.~\ref{sec2}, the oriented areas of Wilson loops are discussed, the numerical results are presented in Sec.~\ref{sec3}, the center vortex explanation of the deviation from the oriented area law observed in the parallel direction case is shown in Sec.~\ref{sec4}, Sec.~\ref{sec5} is a summary.

\section{\label{sec2}Loop Construction}

Center vortices are widely regarded as the dominant degrees of freedom responsible for confinement in $SU(N)$ gauge theory. 
In lattice studies these objects are typically identified after fixing to maximal center gauge and performing center projection. 
In the following discussion the term ``center vortex'' refers to these projected objects unless otherwise stated.
Each vortex that intersects the minimal surface of a Wilson loop contributes a center phase $z\in \mathbb{Z}_3$.
Large, percolating vortex surfaces generate area-law decay of Wilson loops, while small, localized vortices contribute only to short-range correlations.
Neighboring points on a single vortex surface are strongly correlated due to smoothness at the scale of the vortex thickness.

\begin{figure}[htbp]
\includegraphics[width=0.8\hsize]{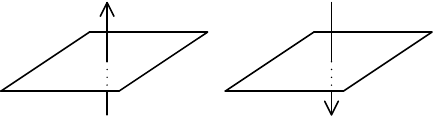}
\caption{\label{fig:oriantedvortex}
When the phase contributed by a center vortex differs in sign, it can be denoted as the center vortex piercing the plane of the Wilson loop from opposite directions. 
Specifically, we can denoted that when the phase contributed by a center vortex piercing the Wilson loop is negative, it corresponds to an exit from the plane containing the Wilson loop~(the left panel); conversely, a positive phase corresponds to an entry into that plane~(the right panel)~\cite{Mickley:2024vkm}.}
\end{figure}
According to the center vortex theory, when considering $\mathbb{Z}_3$, the orientation in which a vortex pierces a plaquette is important.
This is because, in addition to the identity element, the $\mathbb{Z}_3$ group contains two elements, $e^{2\pi i/3}$ and $e^{-2\pi i/3}$. 
As illustrated in Fig.~\ref{fig:oriantedvortex}, We denote vortices contributing the phase $e^{+2\pi i /3}$ by an arrow pointing toward the plane and those contributing $e^{-2\pi i /3}$ by an arrow pointing away from the plane.
This arrow is only a bookkeeping device indicating the sign of the phase contribution.
Thus, the piercing of the Wilson loop by a center vortex can be regarded as directional, which renders the orientation of the Wilson loop meaningful.
It can be expected that, taking directionality into account, the value of a Wilson loop should no longer follow a simple area law but should instead be related to the topology of the Wilson loop.

\begin{figure}[htbp]
\includegraphics[width=0.8\hsize]{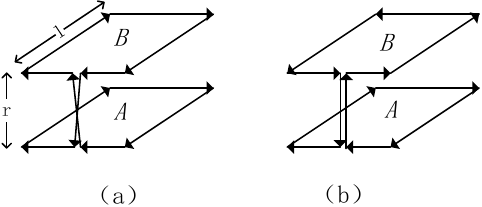}
\caption{\label{fig:fold}
If we define the oriented area enclosed by a Wilson loop within a plane as the area on its right side, then the Wilson loop in the left panel, composed of two small Wilson loops with the same orientation~(denoted as $\langle W_{s.o.}\rangle$), has an oriented area that can be viewed as twice the area of a small plaquette. 
The Wilson loop in the right panel, composed of two small Wilson loops with opposite orientations~(denoted as $\langle W_{o.o.}\rangle$), has an oriented area that can be regarded as zero. 
It is worth noting that the Wilson loop in the left panel is the boundary of a M\"obius strip, while the one in the right panel corresponds to the boundary of an orientable rectangular strip~(a cylindrical strip with a cut).
In the following, the Wilson loops in this figure are denoted as `vertical direction'.}
\end{figure}
\begin{figure}[htbp]
\includegraphics[width=0.8\hsize]{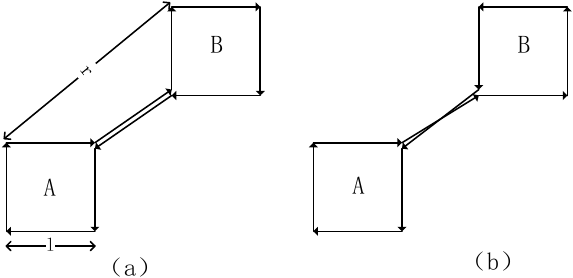}
\caption{\label{fig:paral}
Same as Fig.~\ref{fig:fold} but for case that the two plaquettes are in a same plane.
In the following, the Wilson loops in this figure are denoted as `parallel direction'.}
\end{figure}
Figs.~\ref{fig:fold} and \ref{fig:paral} show the two types of Wilson loops we consider. 
It can be seen that when dealing with Abelian groups, these two Wilson loops correspond to the product of two Wilson loops. 
One advantage of studying these Wilson loops is that they are gauge-invariant, i.e., the results do not depend on the gauge, and therefore no gauge fixing has been applied in our study.
The Wilson loop defined in Figs.~\ref{fig:fold} and \ref{fig:paral} can also serve as a diagnostic of the local coherence and correlation properties of vortex surfaces at varying separations.
While conventional Wilson loops probe the macroscopic effect of many independent vortex piercings, the present observable focuses on correlations, i.e., the interplay of two nearby plaquettes allows one to isolate the effects of local vortex geometry and coherence. 
This provides insight into the transition between single-vortex coherence at short distances and independent random-piercing behavior at larger distances.
A related use of Wilson loops with nontrivial topology was proposed in the study of double-winding Wilson loops in $SU(2)$ gauge theory~\cite{Greensite:2014gra}. 
In that work the dependence of the Wilson loop on the difference of the enclosed areas was shown to favor the center vortex confinement mechanism over alternatives such as the monopole Coulomb gas. 
The Wilson loops considered in the present work have a different geometry, but similarly aim to probe correlations of vortex piercings through surfaces with nontrivial topology.

In the following, the Wilson loops with the contour shown in Fig.~\ref{fig:fold} are denoted as `vertical direction' while those in Fig.~\ref{fig:paral} are denoted as `parallel direction'.
In this work, different $l$ are considered. 
For convenience, all small squares, regardless of size, are referred to as plaquettes.

By using na\"{i}ve area law, one might expect that the two Wilson loops defined in Figs.~\ref{fig:fold} and \ref{fig:paral} would yield identical expectation values since they enclose the same area. 

In the center vortex picture the expectation value of a Wilson loop is determined by the number of vortex piercings of the spanning surface. 
For $SU(3)$, vortices whose associated center phase may take the values $e^{\pm 2\pi i/3}$. 
If the signs of these contributions are correlated, the effective cancellation between opposite phases may depend on the relative orientation of different surface elements. 
Therefore the effective ``area'' controlling the decay of the Wilson loop may deviate from the simple scalar minimal area. 
The Wilson loops considered here provide a probe of such correlations.

Due to the orientation of the plaquettes, a vortex will pierce the Wilson loops in different directions.
If we define the oriented area enclosed by a Wilson loop in a plane as the area on its right side, then the oriented areas of the two are different.
When considering the oriented area of the Wilson loop, in the case of same-orientated~(s.o.) plaquettes as shown in left panels of Figs.~\ref{fig:fold} and \ref{fig:paral}, the area is double that of a single plaquette, while in the case of oppositely-orientated~(o.o.) plaquettes as shown in right panels of Figs.~\ref{fig:fold} and \ref{fig:paral}, the oriented area is zero.
If we simply assume that the expectation value of a Wilson loop still follows the area law, merely with orientation taken into account, then we can obtain the conclusion that, $\langle W _{s.o.} \rangle < \langle W _{o.o.} \rangle$, where $\langle W _{s.o.} \rangle$ and $\langle W _{o.o.} \rangle$ denote the expectation values of the Wilson loops with s.o. and o.o. plaquettes, respectively.
However, the s.o. and o.o. Wilson loops are designed to also capture the nature of non-Abelian gauge which is beyond the center vortex picture.

\section{\label{sec3}Numerical results}

In this work, we study both the cases of quenched approximation and the case with $N_f=2$ dynamical fermions.
When dynamical fermions are included, we employ Kogut-Susskind staggered fermions~\cite{Kogut:1974ag,Kluberg-Stern:1983lmr,Morel:1984di}. 
The full action is given by $S=S_G+S_q$, where
\begin{equation}
\begin{split}
&S_G=\frac{\beta}{N_c}\sum _n \sum _{\mu>\nu}{\rm Retr}\left[1-U_{\mu\nu}(n)\right],\\
&S_q=\sum _n \left(\sum _{\mu}\sum _{\delta = \pm \mu}\bar{\chi}(n) U_{\delta}(n)\eta _{\delta}(n)\chi (n+\delta) \right.\\
&\left.+2am\bar{\chi}\chi \right),
\end{split}
\label{eq.3.1}
\end{equation}
with $a$ being the lattice spacing and $m$ the fermion mass.  
We set $am = 0.1$, which is heavier than the physical mass but light enough for the computational resources available in this exploratory study.  
Here $U_{\mu}=e^{iaA_{\mu}}$, $\eta_{\mu}(n)=(-1)^{\sum_{\nu<\mu}n_{\nu}}$, $U_{-\mu}(n)=U_{\mu}^{\dagger}(n-\mu)$, $\eta_{-\mu}=-\eta_{\mu}(n-\mu)$, and $U_{\mu,\nu}(n)=U_{\mu}(n)U_{\nu}(n+\mu)U_{-\mu}(n+\nu)U_{-\nu}(n)$.  
The simulations are performed using the rational hybrid Monte Carlo algorithm~\cite{Clark:2003na,Clark:2006wp} to implement the ``fourth root trick''.

\subsection{\label{sec3.1}Lattice set up}

\begin{figure}[htbp]
\includegraphics[width=0.49\hsize]{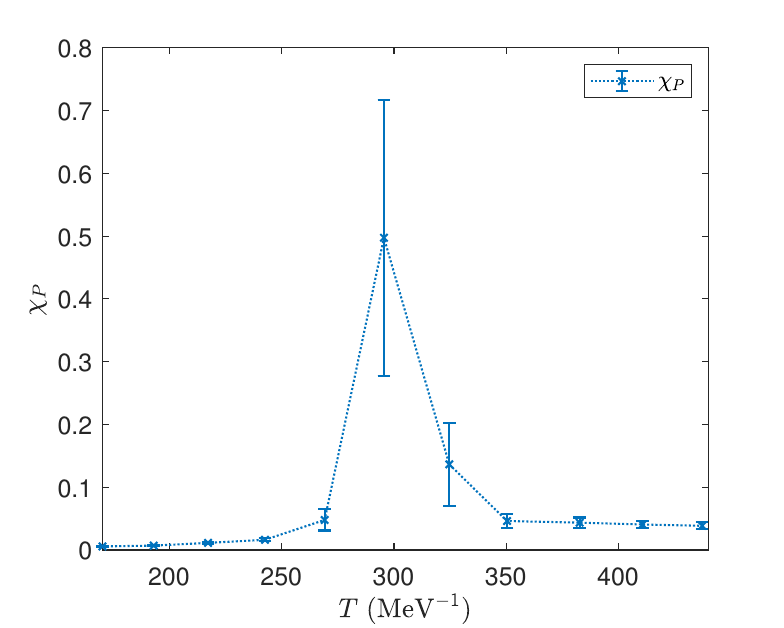}
\includegraphics[width=0.49\hsize]{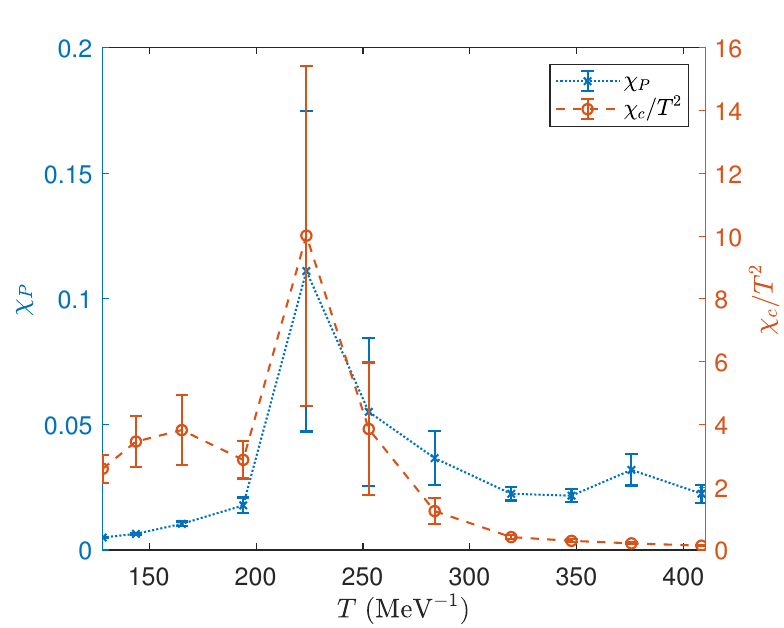}
\caption{\label{fig:polyakovchiral}
$\chi _P$ and $\chi _c$ as functions of temperature in the quenched approximation~(the left panel) and when the dynamical fermions are turned on~(the right panel).}
\end{figure}

\begin{table}
\begin{center}
\begin{tabular}{c|c|c||c|c|c}
\hline    
\multicolumn{3}{c}{Quenched approximation} & \multicolumn{3}{|c}{$N_f=2$ dynamical fermions}  \\
\hline    
$\beta$ & $r_0/a$ & $a^{-1}$ & $\beta$ & $r_0/a$ & $a^{-1}$ \\
& & (MeV) &  &  & (MeV) \\
\hline
$5.65$ & $2.587(9)$ & $1021(4)$ & $5.3$  & $1.951(18)$  & $770(7)$ \\
\hline
$5.7$  & $2.934(6)$ & $1158(2)$ & $5.35$ & $2.187(11)$  & $863(4)$ \\
\hline
$5.75$ & $3.302(5)$ & $1303(2)$ & $5.4$  & $2.513(8)$   & $992(3)$ \\
\hline
$5.8$  & $3.688(5)$ & $1455(2)$ & $5.45$ & $2.949(5)$   & $1164(2)$ \\
\hline
$5.85$ & $4.092(4)$ & $1615(1)$ & $5.5$  & $3.398(5)$   & $1341(2)$ \\
\hline
$5.9$  & $4.493(4)$ & $1773(2)$ & $5.55$ & $3.841(4)$   & $1516(2)$ \\
\hline
$5.95$ & $4.935(4)$ & $1948(2)$ & $5.6$  & $4.310(4)$   & $1701(2)$ \\
\hline
$6.0$  & $5.326(4)$ & $2102(2)$ & $5.65$ & $4.854(4)$   & $1916(2)$ \\
\hline
$6.05$ & $5.818(5)$ & $2296(2)$ & $5.7$  & $5.283(5)$   & $2085(2)$ \\
\hline
$6.1$  & $6.244(6)$ & $2464(2)$ & $5.75$ & $5.709(5)$   & $2253(2)$ \\
\hline
$6.15$ & $6.648(6)$ & $2623(2)$ & $5.8$  & $6.207(6)$   & $2450(2)$ \\
\hline
\end{tabular}
\end{center}
\caption{\label{tab.matching}Results of matching for different values of $\beta$.}
\end{table}
To determine the temperatures corresponding to different values of $\beta$, we perform a matching on an $L_x \times L_y \times L_z \times L_\tau = 24^3 \times 48$ lattice, where $L_{x,y,z,\tau}$ denote the extents in the respective directions.
In the quenched approximation, simulations are carried out for $\beta = 5.65, 5.7, 5.75, \ldots, 6.15$. 
For the case with dynamical fermions ($N_f=2$), simulations are performed at $\beta = 5.3, 5.35, 5.4, \ldots, 5.8$.
For each $\beta$ value, a total of $100 + 900$ molecular dynamics time units (TUs) are generated. 
The first $100$ TUs are discarded for thermalization, and the subsequent $900$ TUs are used for measurements.
The lattice spacing $a$ is determined by calculating the static quark potential $V(r)$~\cite{Bali:1992ab,Bali:2000vr,Orth:2005kq} and matching the resulting `Sommer scale' $r_0$ to its physical value of $0.5\;\text{fm}$~\cite{Sommer:1993ce,Cheng:2007jq,MILC:2010hzw}.
The results of lattice spacings are shown in Table~\ref{tab.matching}.
Throughout this work, we estimate the statistical error using $\sigma = \sqrt{2 \tau _{\rm int}} \sigma _{\rm jk}$~\cite{Gattringer:2010zz}. 
Here, $\tau _{\rm int}$ denotes the separation in TUs required for two configurations to be considered statistically independent, and $\sigma _{\rm jk}$ is the error computed via the `jackknife method'. 
The autocorrelation time $\tau _{\rm int}$ is determined using an autocorrelation analysis with a window parameter $S = 1.5$~\cite{Wolff:2003sm}. 

In this work, we present the results for Wilson loops measured on a lattice of size $24^3 \times 6$. 
When $L_t=6$, for each $\beta$, a total of $100 + 2900$ TUs are generated, where first $100$ TUs are discarded for thermalization, and the subsequent $2900$ TUs are used for measurements..
For the parameters listed in Table~\ref{tab.matching}, the corresponding phases at $L_{\tau} = 6$ can be identified via the Polyakov loop and the chiral condensate.
The quantities are defined as~\cite{Bazavov:2011nk},
\begin{equation}
\begin{split}
&P=\frac{1}{N_cL_xL_yL_z}\sum _{\vec{n}} \prod _{n_{\tau}}U_{\tau}(\vec{n}),\\
&\chi _P = a^3L_xL_yL_z\left(\langle P ^2 \rangle - \langle P\rangle ^2\right),\\
&\chi _{c}=\frac{1}{a^2L_xL_yL_zL_{\tau}}\left(\langle {\rm tr}\left[D^{-1}\right]^2\rangle-\langle {\rm tr}\left[D^{-1}\right]\rangle^2\right),\\
\end{split}
\label{eq.cqdefine}
\end{equation}
where $P$ is the bare Polyakov loop, $\chi _P$ is the susceptibility of $P$, and $\chi _{c}$ is the disconnected susceptibility of the bare chiral condensation.
In the case of the quenched approximation, we use $\chi _{P} =  a^3L_xL_yL_z \left(\langle\left| P\right| ^2 \rangle - \langle \left|P\right|\rangle ^2\right)$.
The results of $\chi _{P}$ and $\chi _c$ are shown in Fig.~\ref{fig:polyakovchiral}.
It can be seen that, the parameters in consideration cover the confined/deconfined phases.

\subsection{\label{sec3.3}Correlation of two Wilson loops in the vertical direction}

We now consider two parallel planes, separated by a distance $r$ along the $z$-direction. 
In each plane, a square Wilson loop of side length $l$ (where $l$ ranges from $1$ to $6$) is placed such that their projections onto the $x-y$ plane are identical and aligned. 
The distance $r$ is measured in lattice units and varies from $1$ to $12$.

To construct a closed Wilson loop that connects these two squares across the planes, we generate two identical square loops in their respective planes and link them with vertical segments. 
The vertical connection is chosen to be a straight line along the $z$-direction, connecting the midpoint of the bottom edge of each square. 
Specifically, we define a reference point at coordinates $(l/2, 0)$ relative to each square's bottom-left corner. 
This point serves as both the endpoint of the square loop in plane $A$ and the starting point of the square loop in plane $B$, ensuring that the vertical link is a direct, perpendicular connection between the two planes at this corresponding location.

The closed loop is then constructed by concatenating four segments in order, which are
\begin{enumerate}
\item The square loop in plane $A$ (traversed counterclockwise), starting and ending at the midpoint of its bottom edge.
\item A vertical connection from plane $A$ to plane $B$, consisting of $r$ consecutive upward steps.
\item The square loop in plane $B$, which can be traversed either clockwise or counterclockwise, depending on whether $W_{o.o}$ or $W_{s.o.}$ is measured, also starting and ending at the midpoint of its bottom edge.
\item A vertical connection from plane $B$ back to plane $A$, consisting of $r$ consecutive downward steps.
\end{enumerate}
This systematic generation provides a complete set of Wilson loop paths spanning two parallel planes, suitable for studying inter-plane correlations or computing expectation values in lattice gauge theory simulations.
Similarly, an average over all possible spatial directions and starting points on the lattice is performed to obtain the measured value $W(r)$ for one gauge field configuration.

\begin{figure}[htbp]
\includegraphics[width=0.49\hsize]{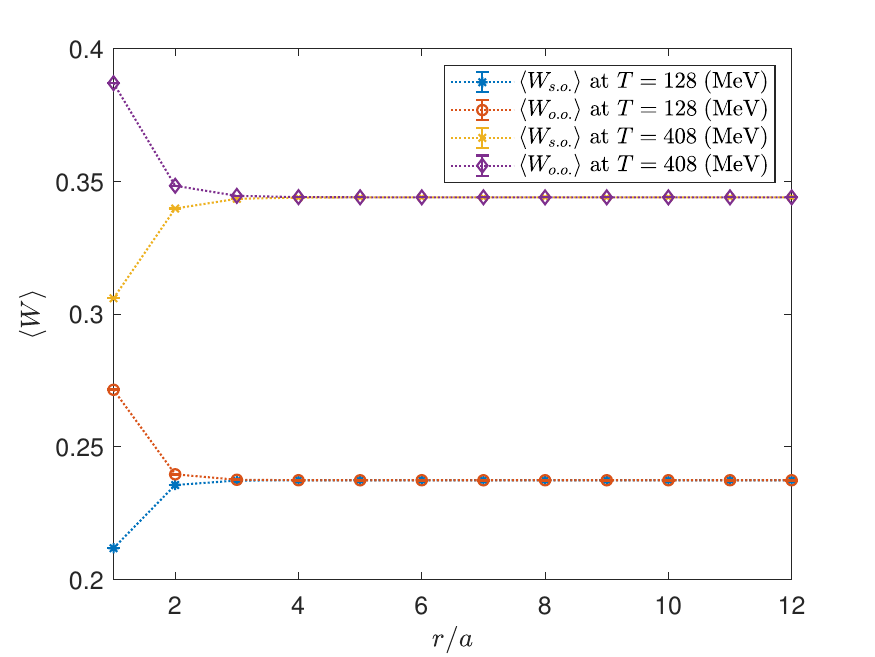}
\includegraphics[width=0.49\hsize]{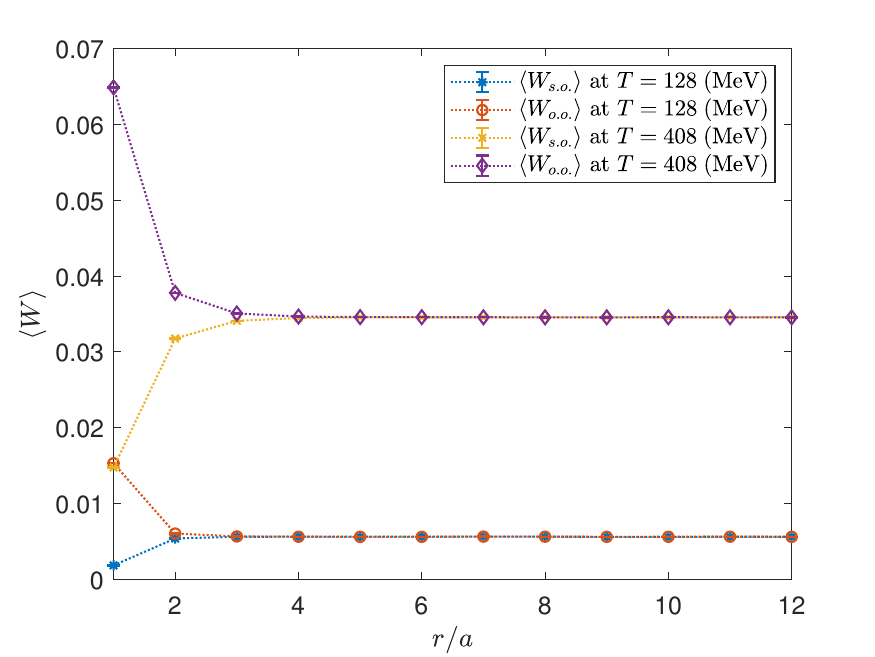}\\
\includegraphics[width=0.49\hsize]{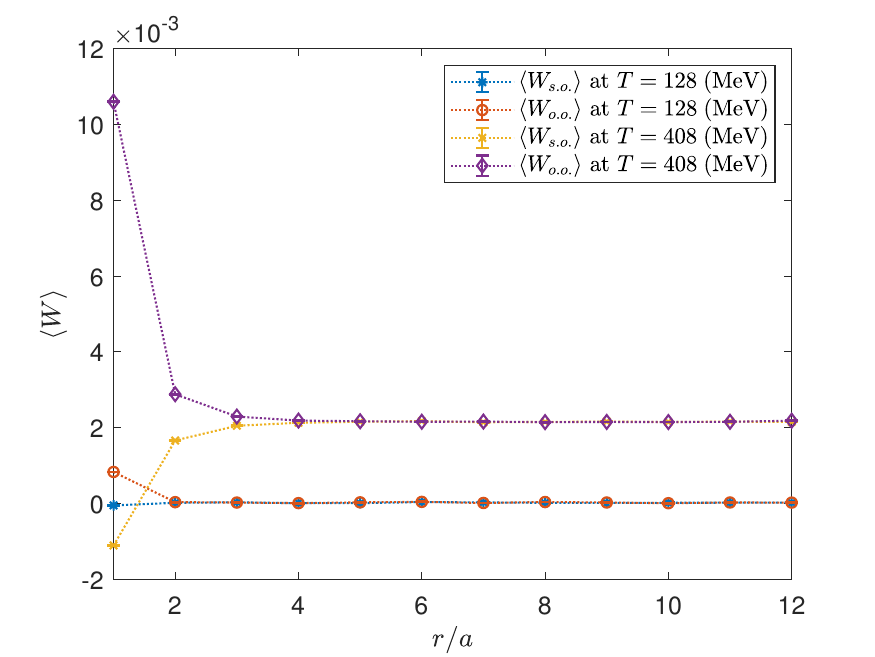}
\includegraphics[width=0.49\hsize]{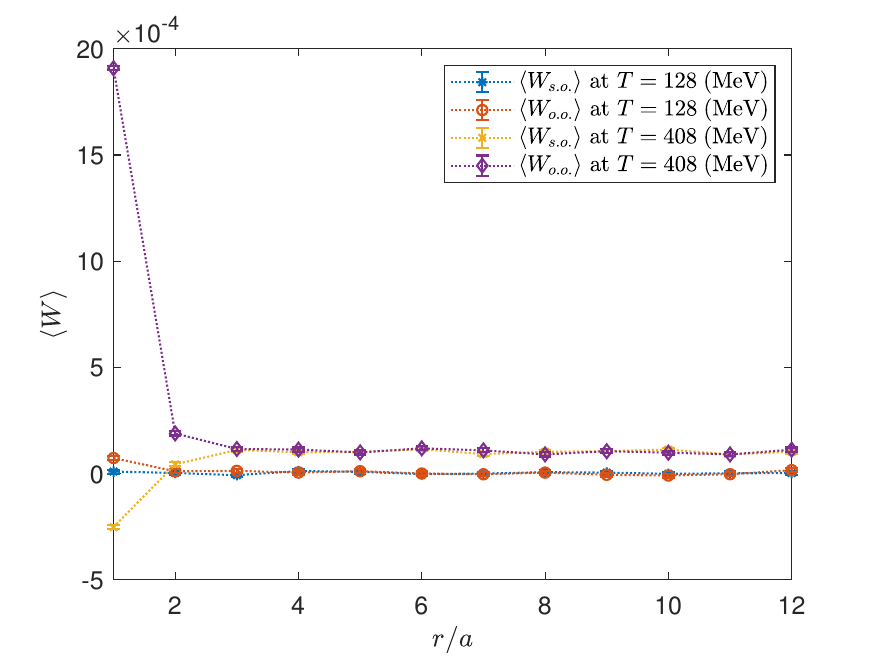}
\caption{\label{fig:bs01f}
$\langle W_{o.o}\rangle$ and $\langle W_{s.o.}\rangle$ in the case of vertical direction~(as shown in Fig.~\ref{fig:fold}) at high and low temperatures in quenched approximation.
The top-left panel corresponds to $l=1$, the top-right panel corresponds to $l=2$, The bottom-left panel corresponds to $l=3$, and the bottom-right panel corresponds to $l=4$, respectively.}
\end{figure}
\begin{figure}[htbp]
\includegraphics[width=0.49\hsize]{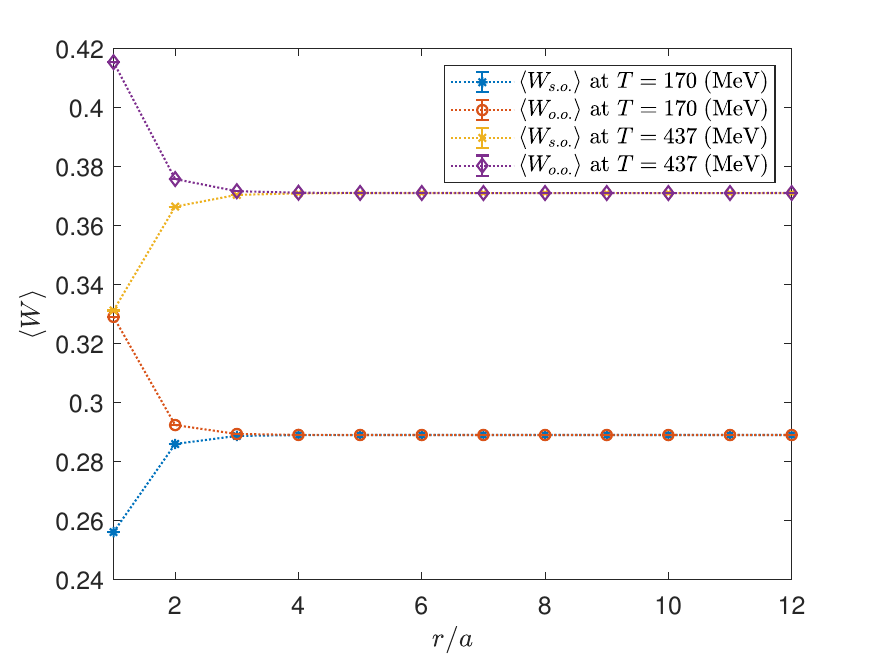}
\includegraphics[width=0.49\hsize]{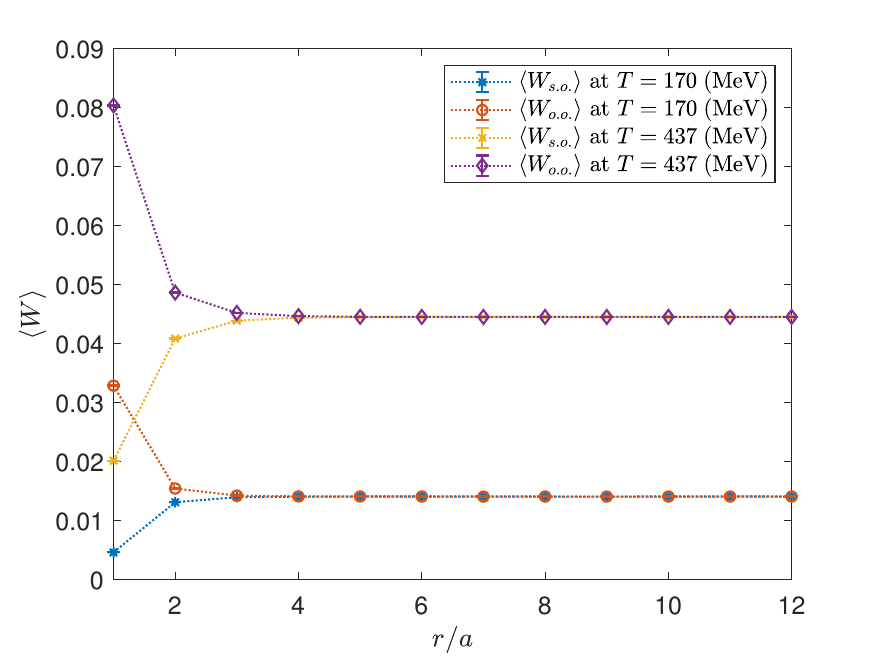}\\
\includegraphics[width=0.49\hsize]{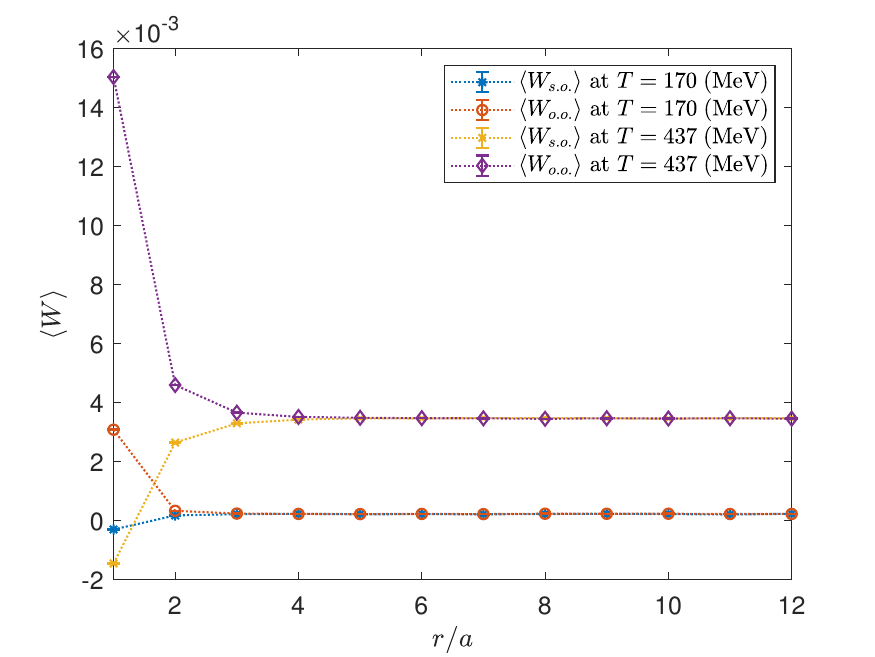}
\includegraphics[width=0.49\hsize]{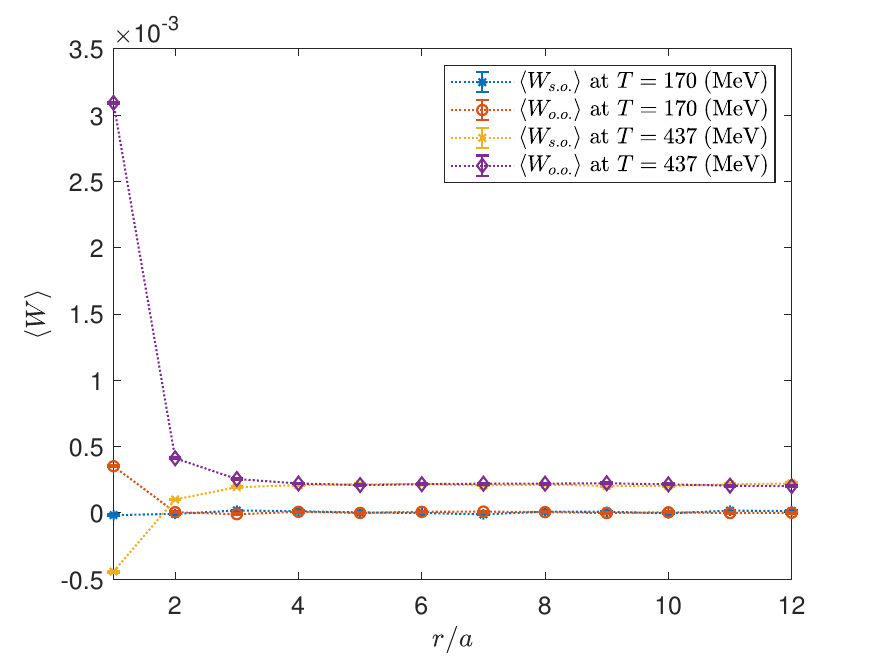}
\caption{\label{fig:bqf}Same as Fig.~\ref{fig:bs01f} but for the case when the dynamical fermions are turned on.}
\end{figure}
The results of $\langle W_{o.o}(r) \rangle$ and $\langle W_{s.o.}(r) \rangle$ in both the quenched approximation and with dynamical fermions are shown in Figs.~\ref{fig:bs01f} and \ref{fig:bqf}, respectively.
It can be observed that, at large $r$, $\langle W_{o.o}(r) \rangle$ and $\langle W_{s.o.}(r) \rangle$ approach each other, which indicates that the orientation of the two plaquettes becomes less relevant at large separations.
However, at small $r$, a significant difference exists between $\langle W_{o.o}(r) \rangle$ and $\langle W_{s.o.}(r) \rangle$.
As expected, $\langle W_{s.o.}(r) \rangle < \langle W_{o.o}(r) \rangle$ is found.

\begin{figure}[htbp]
\includegraphics[width=0.49\hsize]{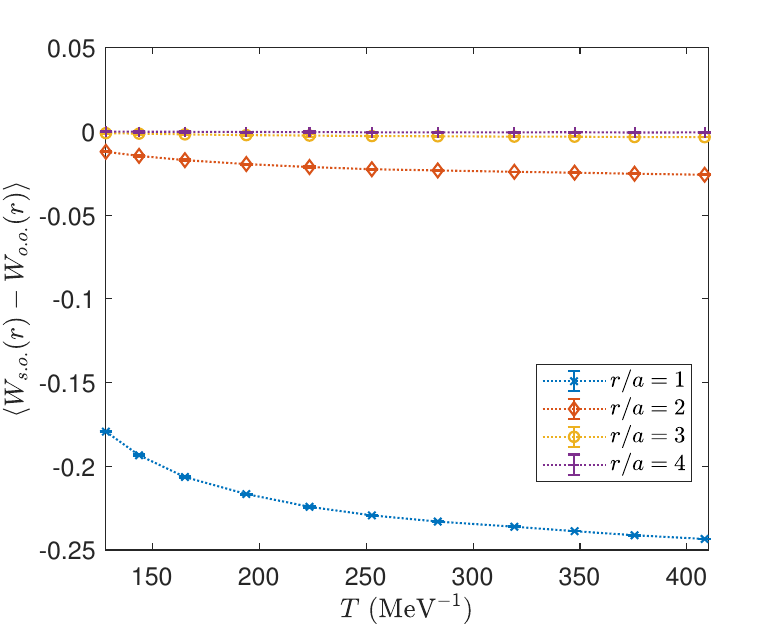}
\includegraphics[width=0.49\hsize]{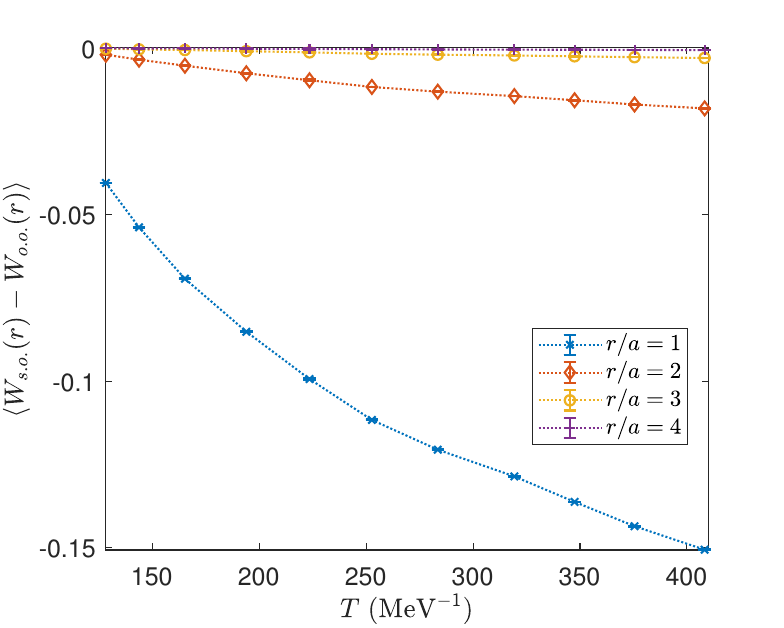}\\
\includegraphics[width=0.49\hsize]{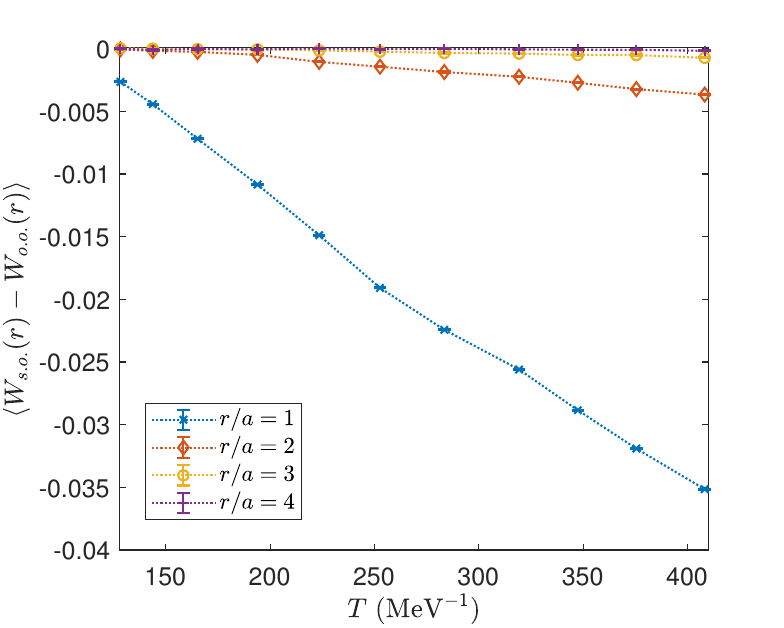}
\includegraphics[width=0.49\hsize]{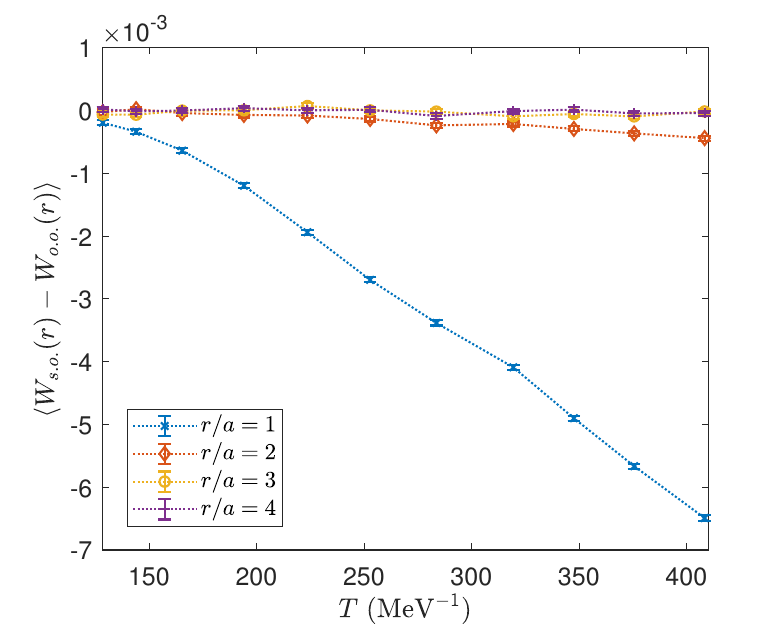}
\caption{\label{fig:bs01ft}
The differences between $\langle W_{s.o.}\rangle $ and $\langle W_{o.o.}\rangle$ in the case of vertical direction at different temperatures in quenched approximation.
The top-left panel corresponds to $l=1$, the top-right panel corresponds to $l=2$, The bottom-left panel corresponds to $l=3$, and the bottom-right panel corresponds to $l=4$, respectively.
It should be noted that the bare Wilson loops are measured, so we cannot directly compare Wilson loops at different temperatures.
}
\end{figure}
\begin{figure}[htbp]
\includegraphics[width=0.49\hsize]{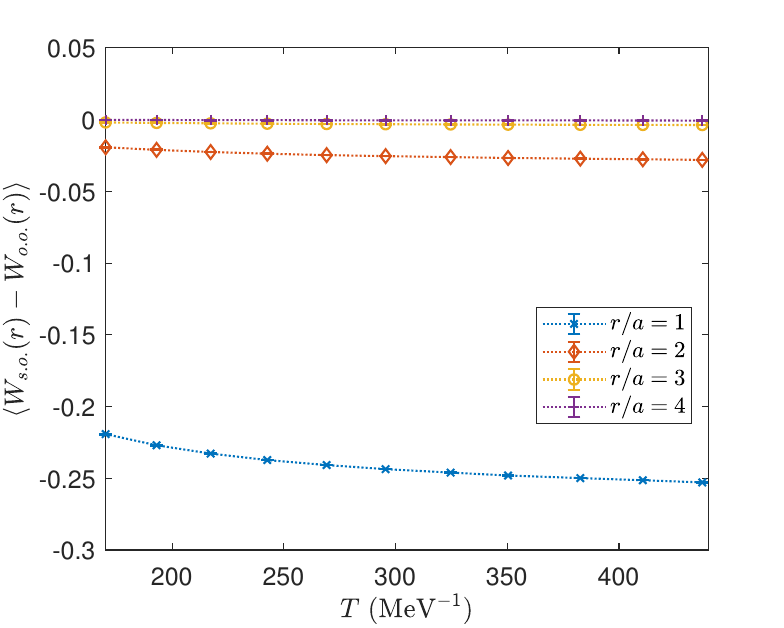}
\includegraphics[width=0.49\hsize]{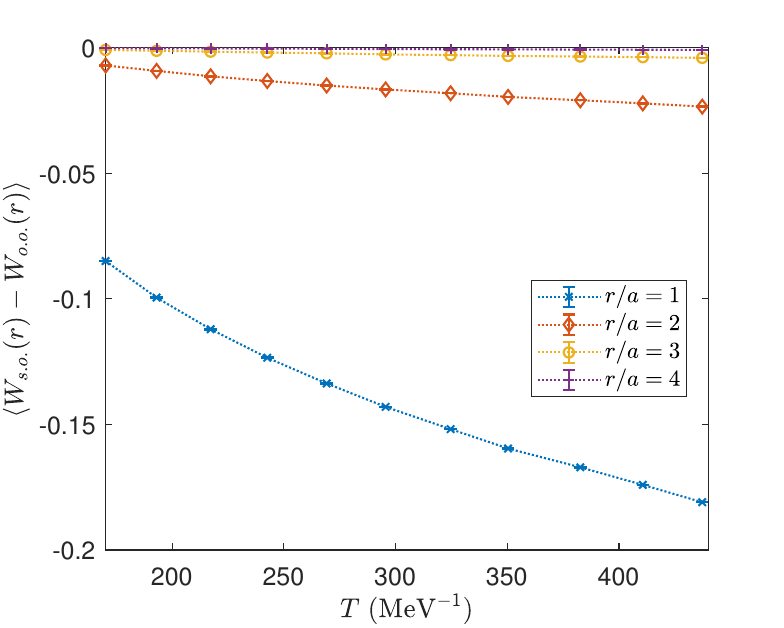}\\
\includegraphics[width=0.49\hsize]{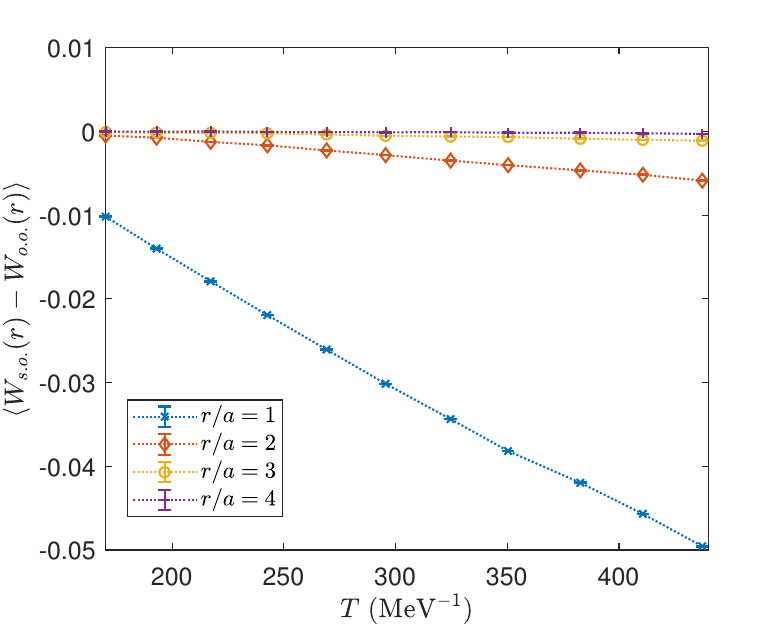}
\includegraphics[width=0.49\hsize]{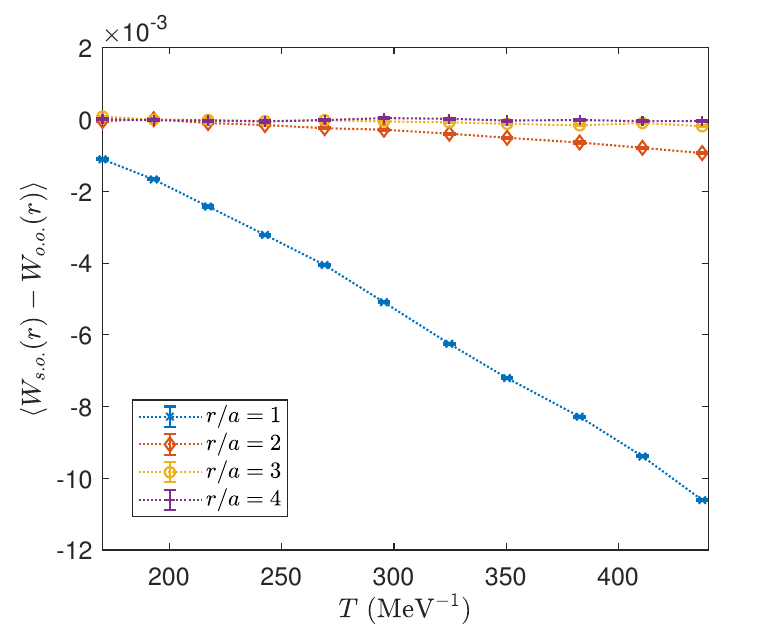}
\caption{\label{fig:bqft}
Same as Fig.~\ref{fig:bs01ft} but for the cases when the dynamical fermions are turned on.}
\end{figure}

The differences between $\langle W_{s.o.}\rangle $ and $\langle W_{o.o.}\rangle $ at different temperatures are shown in Figs.~\ref{fig:bs01ft} and \ref{fig:bqft}.
It should be noted that the bare Wilson loops are measured, so we cannot directly compare Wilson loops under different parameters~(i.e., at different temperatures). 
Figs.~\ref{fig:bs01ft} and \ref{fig:bqft} merely show that differences in the $\langle W_{s.o.}\rangle $ and $\langle W_{o.o.}\rangle $ can occur at various temperatures, whether in the confinement phase or the deconfinement phase.

\subsection{\label{sec3.2}Correlation of two Wilson loops in the parallel direction}

We consider a planar Wilson loop as shown in Figure 1, which consists of two separate square regions, denoted as $A$ and $B$, with equal side length $l$ (ranging from $1$ to $6$). 
The distance between the centers of the two squares is denoted by $r$, and we examine all cases where $l \leq r \leq 12$. 
We only consider cases where square $B$ is located to the upper right of square $A$, i.e., its bottom-left corner has coordinates $(x, y)$ with $x \geq 0, y \geq 0$, and at least one of $x \geq l$ or $y \geq l$ holds, ensuring that the two squares do not overlap.

The Wilson link connecting the two squares is represented by a straight diagonal line. 
The diagonal is chosen as the shortest possible line connecting the boundaries of square $A$ and square $B$. 
When multiple shortest lines exist (e.g., when $B$ is directly above or to the right of $A$), we choose the line that is closest to the line connecting the centers of the two squares (i.e., the more central one).

The diagonal line does not lie on the lattice grid. 
To approximate this on the lattice, we replace the diagonal with a piecewise linear path that intersects it. 
To do this, we employ an algorithm similar to Bresenham's line algorithm. 
The algorithm constructs a path from a starting point $(0,0)$ to an endpoint $(dx, dy)$, where $dx$ and $dy$ are non-negative integers representing the horizontal and vertical distances between the two connecting points. 
At each step, the algorithm evaluates which of the two possible moves, right (increase $x$ by 1) or up (increase $y$ by 1), will keep the path closer to the ideal diagonal. 
This is done by comparing the perpendicular distances from the points $(x+1, y)$ and $(x, y+1)$ to the ideal line. 
The move that yields the smaller distance is chosen. 
If the distances are equal, the right move is selected. 
If one of the directions has already reached its maximum ($x = dx$ or $y = dy$), the remaining steps are taken in the other direction until the endpoint is reached.

The following is a summary of the implementation of the Wilson loop path construction,
\begin{enumerate}
\item For each valid position of square B, we first determine the connecting points on squares A and B as described above. We then generate four path segments.
\item A closed loop around square $A$ (clockwise) is generated.  
\item A connecting path from square $A$ to square $B$ using the Bresenham-like algorithm is generated.  
\item A closed loop around square $B$, which can be either clockwise or counterclockwise depending on whether $W_{o.o}$ or $W_{s.o.}$ is measured.  
\item A return connecting path from square $B$ back to square $A$, which is the reverse of the forward connecting path.  
\item These segments are concatenated to form a complete, closed Wilson loop.
\end{enumerate}

After constructing a Wilson loop path, the actual measurements are performed by replacing the $x-y$ directions with $x-y$, $x-z$, and $y-z$ respectively, and then letting the starting point of the Wilson loop traverse all lattice sites. 
For a given gauge field configuration, the average over all these cases is taken as the measured value for that configuration. 
Subsequently, for the same configuration, an average over all Wilson loop paths with the same $r$ is calculated, and this is treated as the measured value of $W(r)$ for that configuration. 
The final expectation value $\langle W(r) \rangle$ is obtained by averaging $W(r)$ over all gauge configurations.

\begin{figure}[htbp]
\includegraphics[width=0.49\hsize]{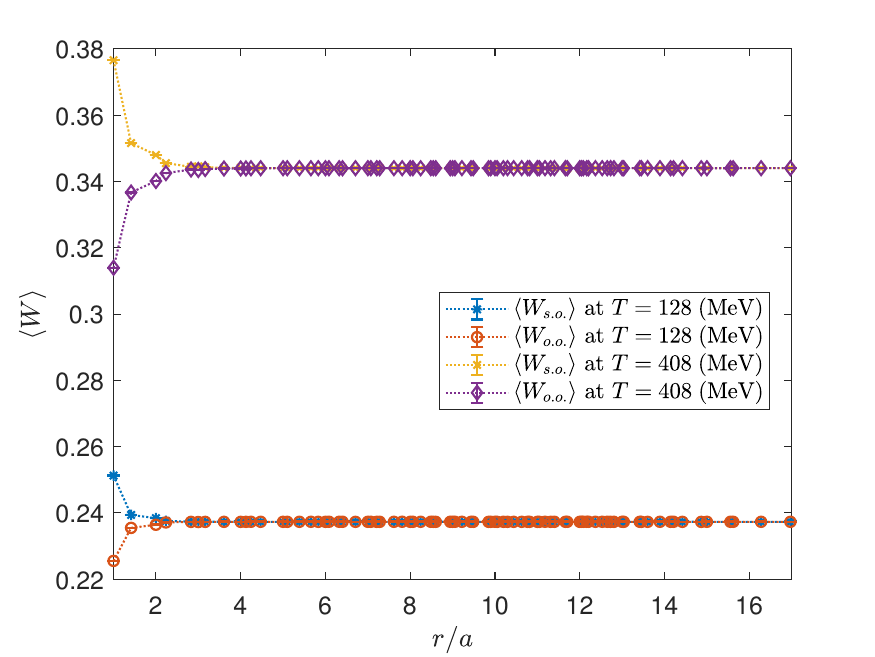}
\includegraphics[width=0.49\hsize]{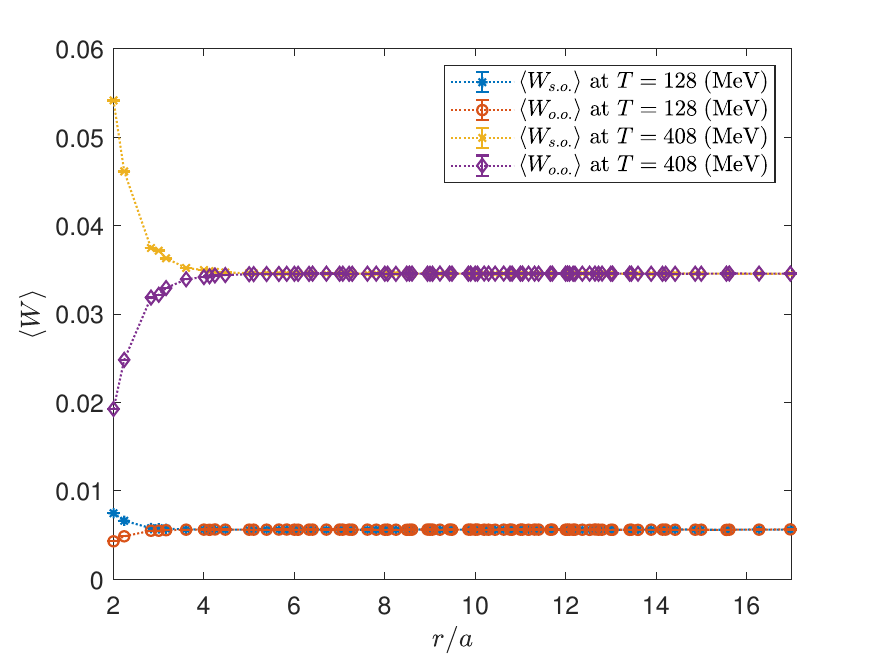}\\
\includegraphics[width=0.49\hsize]{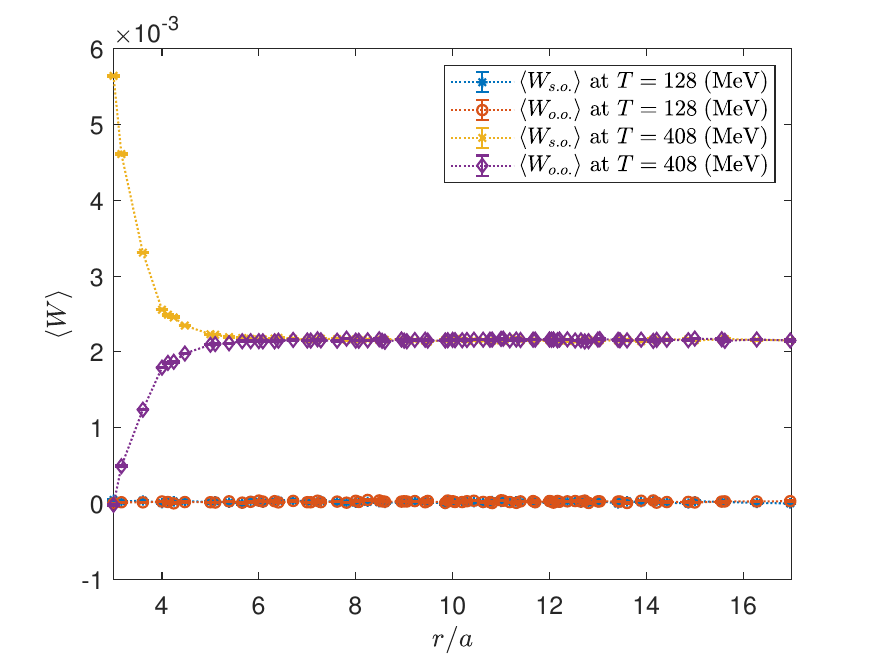}
\includegraphics[width=0.49\hsize]{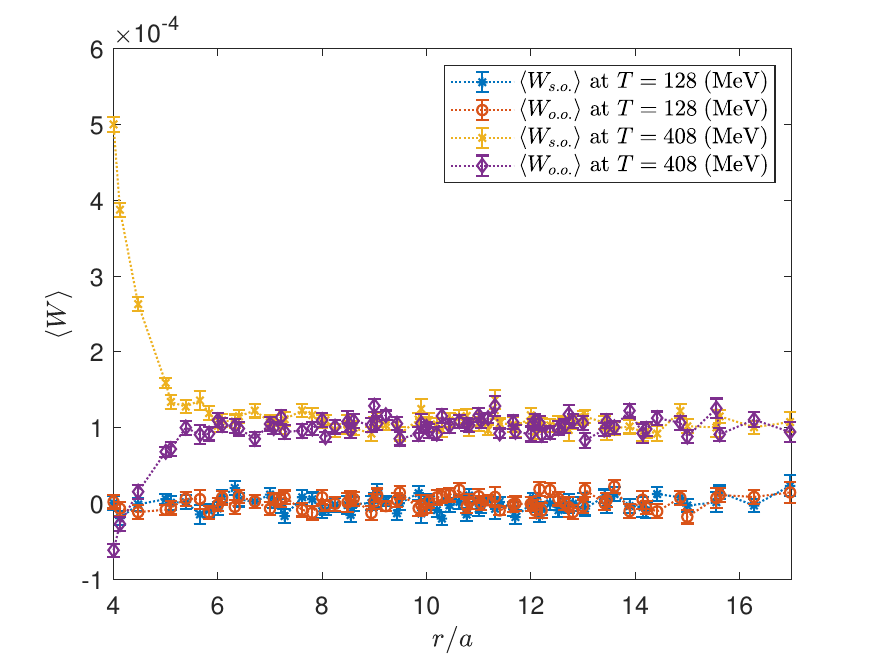}
\caption{\label{fig:bs01k}
Same as Fig.~\ref{fig:bs01f} but for the case of parallel direction.}
\end{figure}
\begin{figure}[htbp]
\includegraphics[width=0.49\hsize]{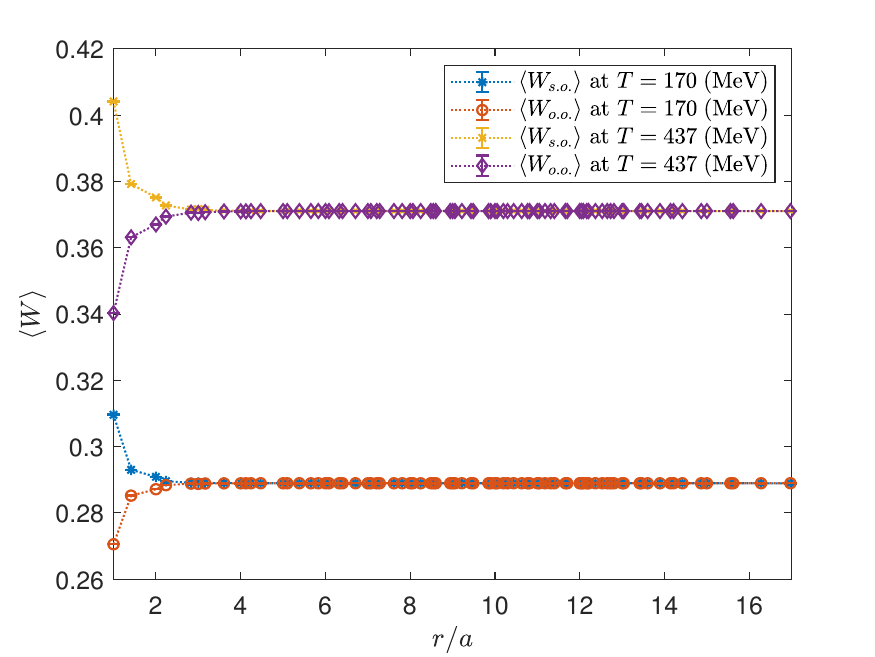}
\includegraphics[width=0.49\hsize]{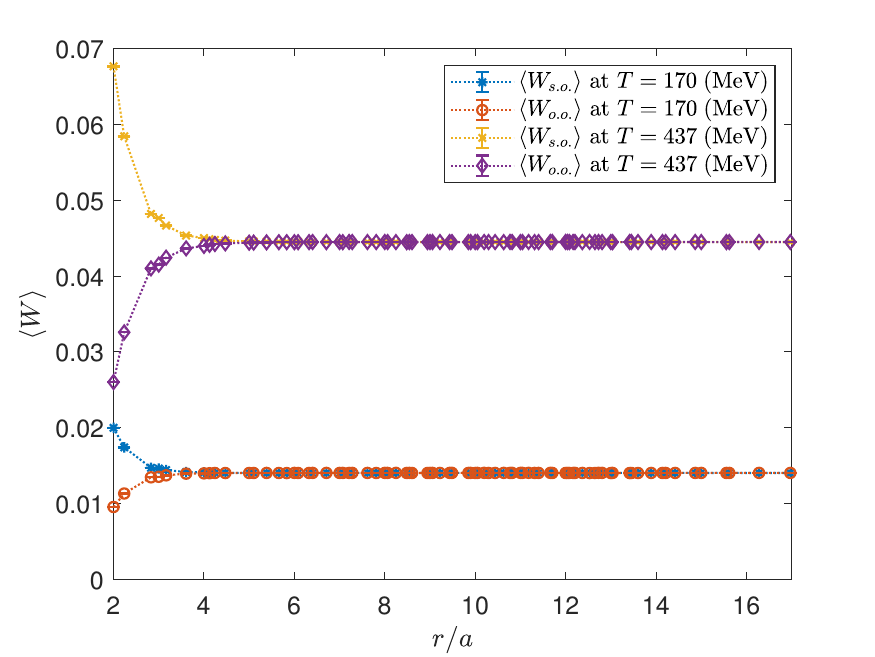}\\
\includegraphics[width=0.49\hsize]{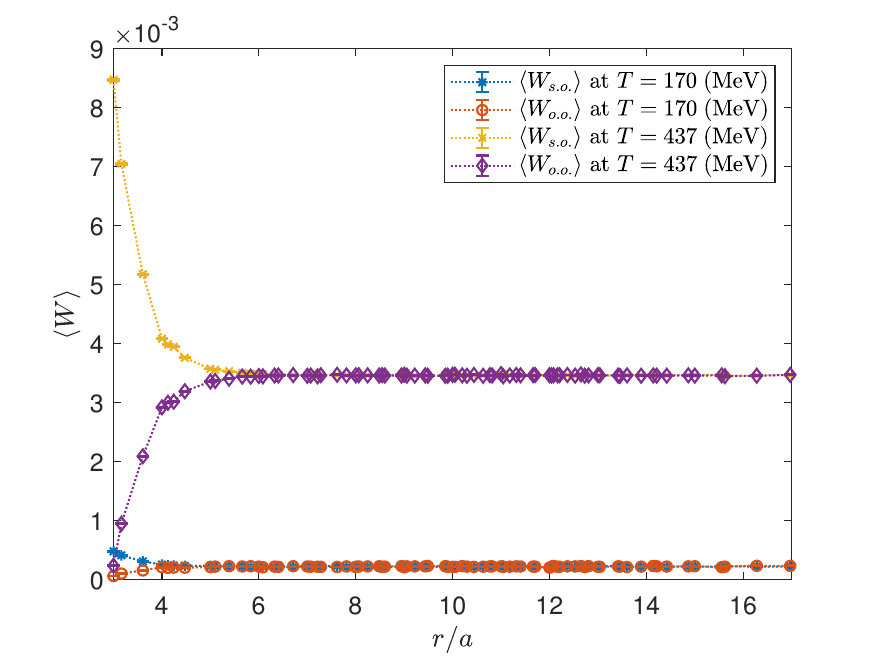}
\includegraphics[width=0.49\hsize]{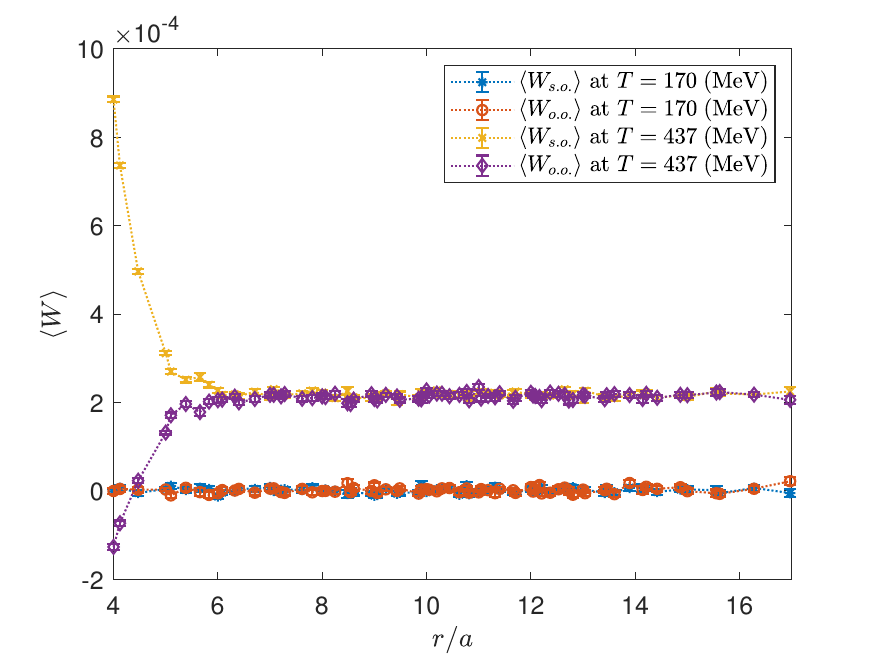}
\caption{\label{fig:bsqk}
Same as Fig.~\ref{fig:bs01k} but for the cases when the dynamical fermions are turned on.}
\end{figure}
The results of $\langle W_{o.o}(r) \rangle$ and $\langle W_{s.o.}(r) \rangle$ are shown in Figs.~\ref{fig:bs01k} and \ref{fig:bs01k}.
Similar to the case where the loops are separated in the vertical direction, a difference in $\langle W_{o.o}(r) \rangle$ and $\langle W_{s.o.}(r) \rangle$ is also observed which decays with growing $r$.
However, contrary to the naive expectation based on oriented area, we find that $\langle W_{s.o.}(r) \rangle > \langle W_{o.o}(r) \rangle$ at small $r$.
Apart from that, for large plaquettes, the difference becomes smaller in confined phase, and cannot be observed when $l$ is large~(when $l=3,4$ in quenched approximation, and when $l=4$ with dynamical fermions turned on).

\begin{figure}[htbp]
\includegraphics[width=0.49\hsize]{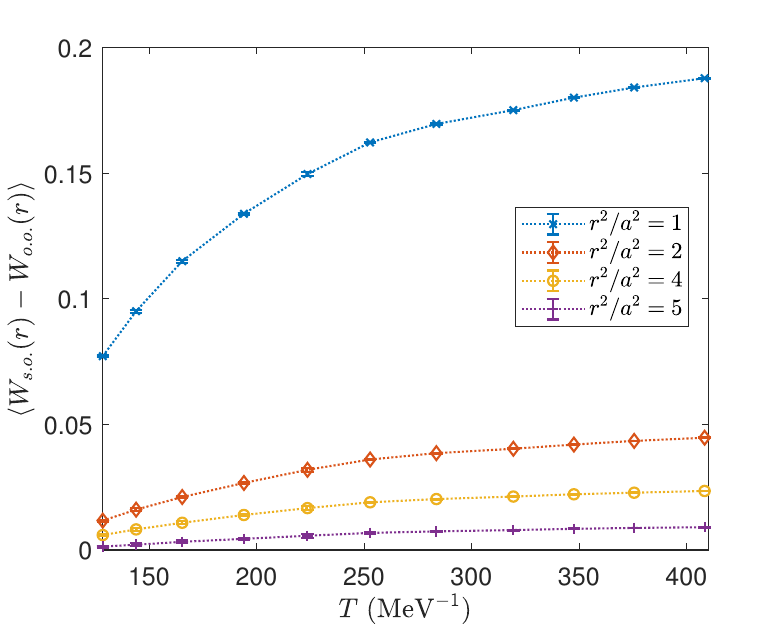}
\includegraphics[width=0.49\hsize]{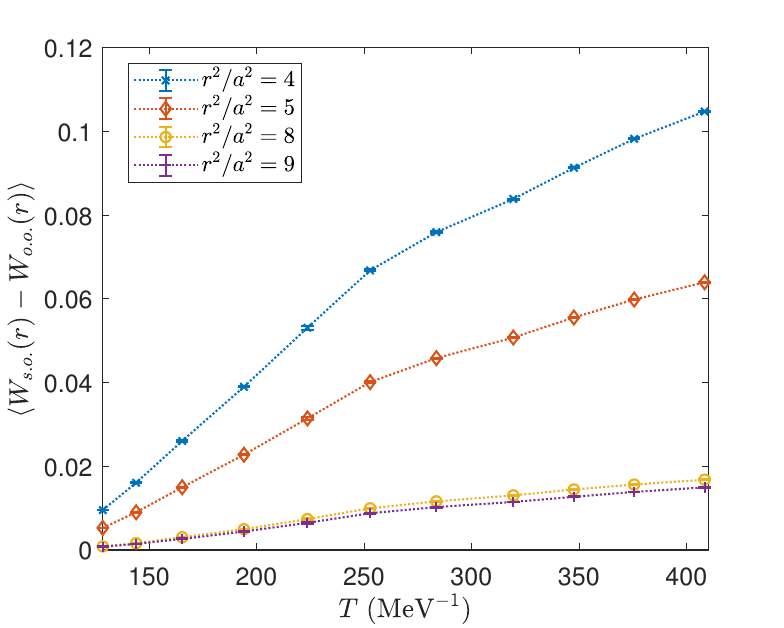}\\
\includegraphics[width=0.49\hsize]{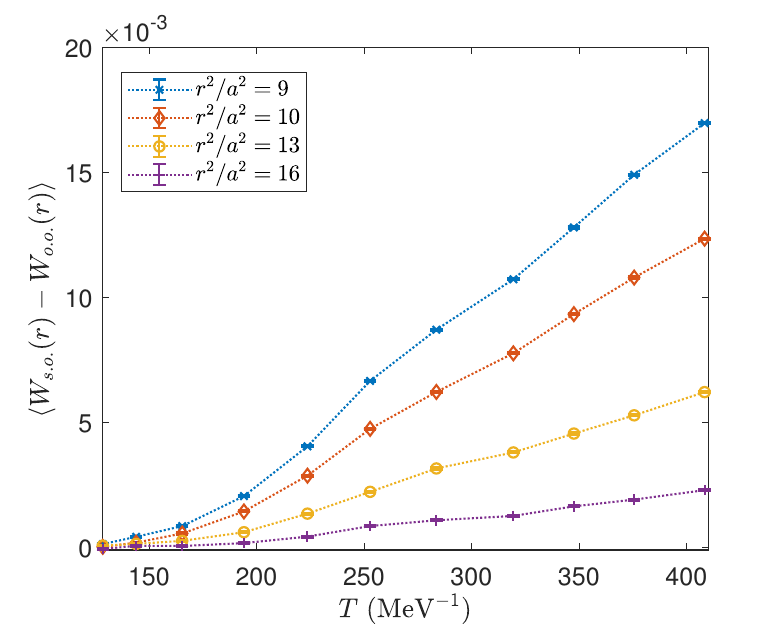}
\includegraphics[width=0.49\hsize]{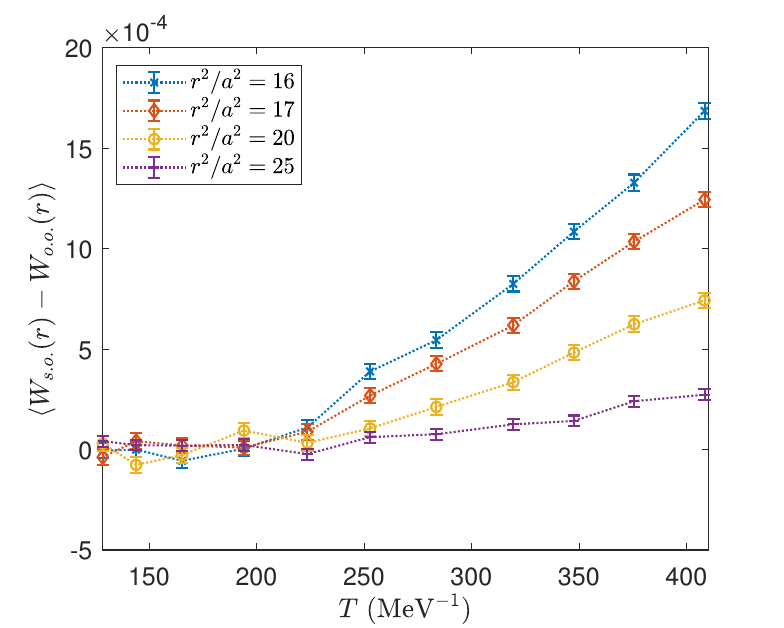}
\caption{\label{fig:bs01kt}
Same as Fig.~\ref{fig:bs01ft} but for the case of parallel direction.}
\end{figure}
\begin{figure}[htbp]
\includegraphics[width=0.49\hsize]{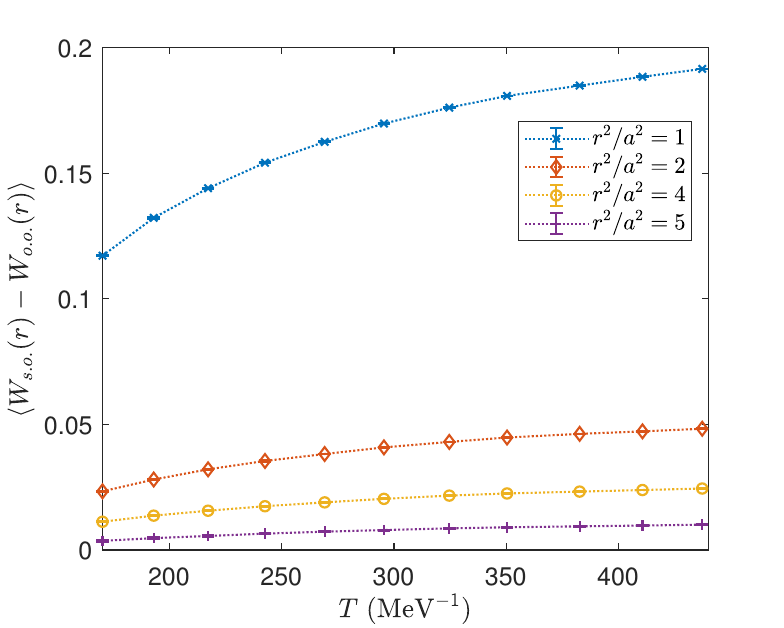}
\includegraphics[width=0.49\hsize]{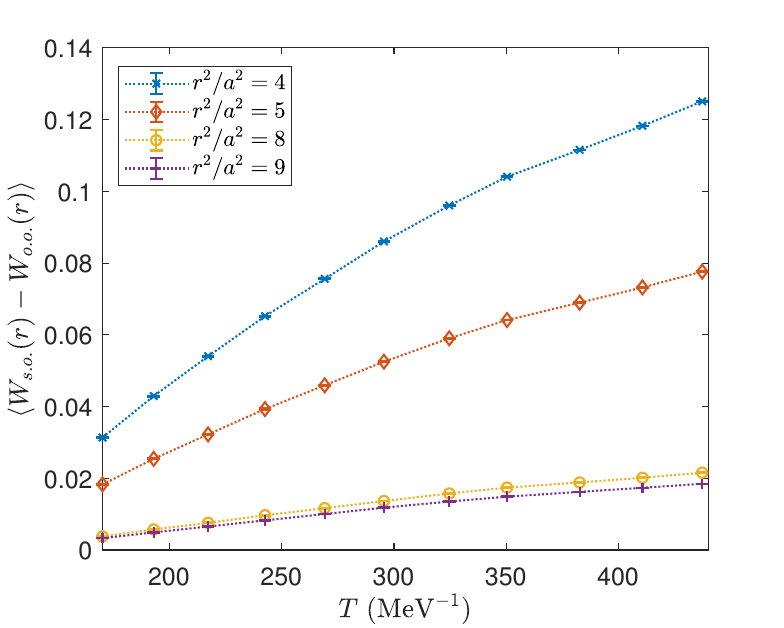}\\
\includegraphics[width=0.49\hsize]{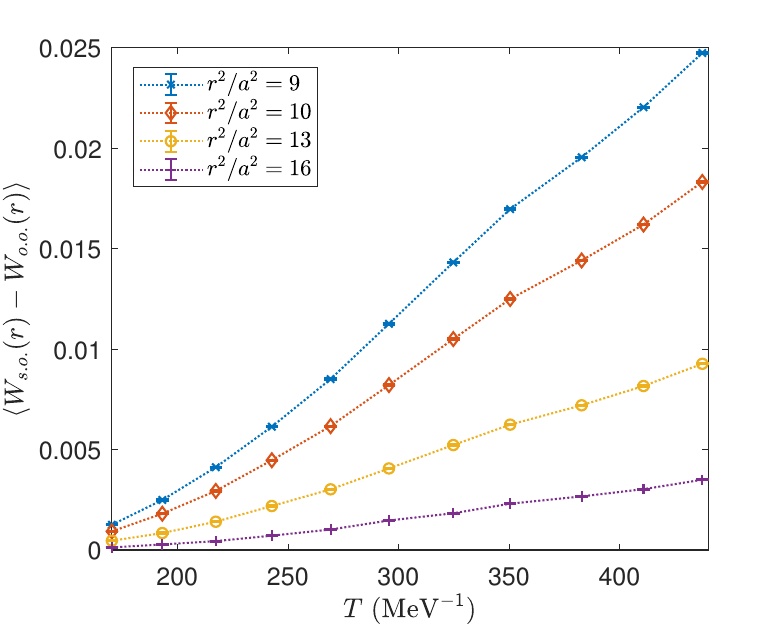}
\includegraphics[width=0.49\hsize]{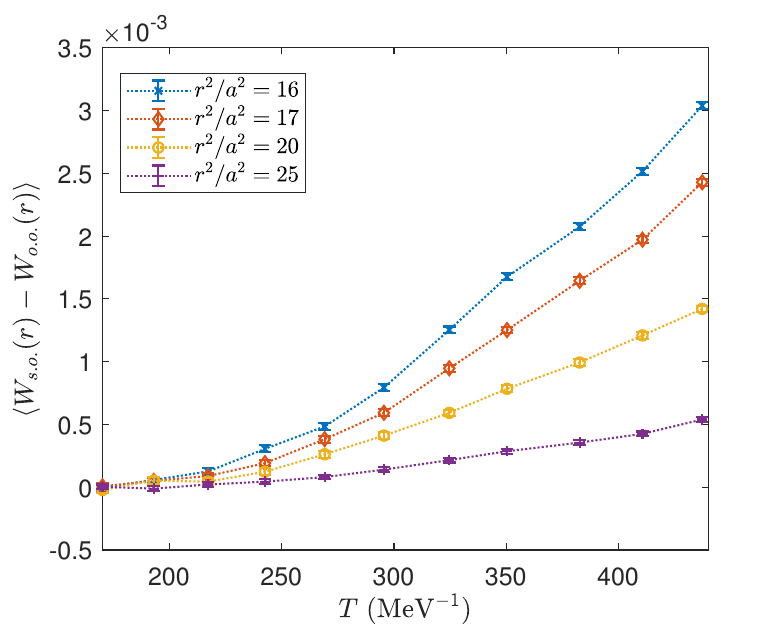}
\caption{\label{fig:bsqkt}
Same as Fig.~\ref{fig:bs01kt} but for the cases when the dynamical fermions are turned on.}
\end{figure}

The differences between $\langle W_{s.o.}\rangle $ and $\langle W_{o.o.}\rangle $ at different temperatures are shown in Figs.~\ref{fig:bs01kt} and \ref{fig:bsqkt}.
Similarly as the previous subsection, it should be noted that the bare Wilson loops are measured so that one cannot directly compare the results at different temperatures.
However, it can be shown that the differences in the $\langle W_{s.o.}\rangle $ and $\langle W_{o.o.}\rangle $ can occur at various temperatures.

\section{\label{sec4}Center vortex explanation of the failure of a naive oriented area rule.}

\begin{figure}[htbp]
\includegraphics[width=0.8\hsize]{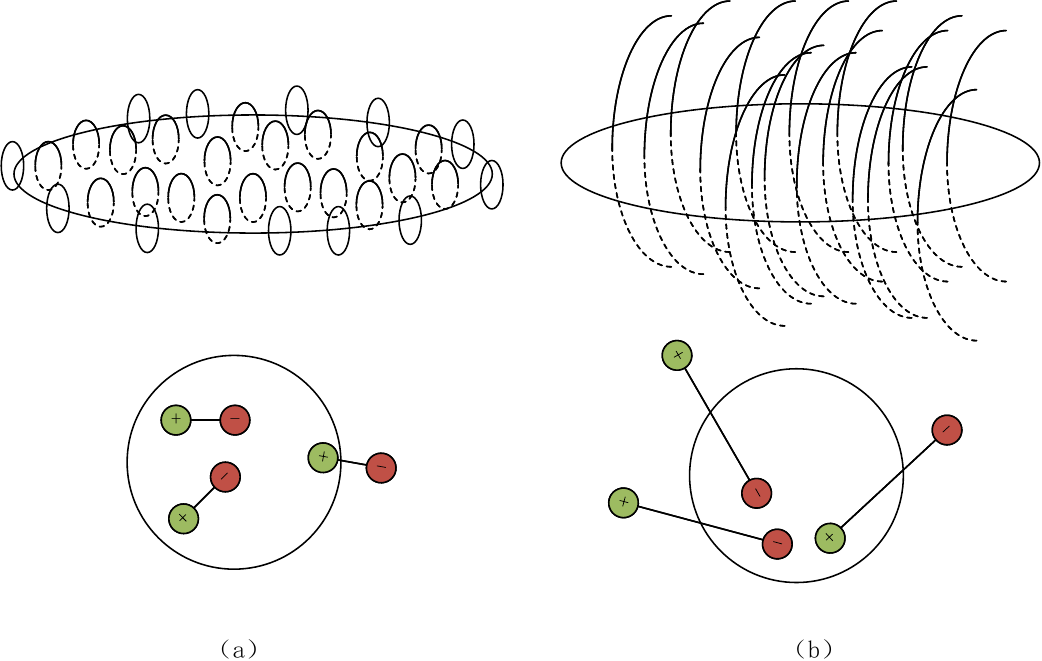}
\caption{\label{fig:percolationpicture}
Schematic of percolation of center vortices in three-dimensional space. 
In three-dimensional space, center vortices are closed curves. 
The left side of the first row illustrates that when the scale of center vortices is small, those capable of penetrating the Wilson loop an odd number of times are distributed near the boundary of the Wilson loop. 
The right side shows that when the scale of center vortices is large, the number of center vortices capable of penetrating the Wilson loop an odd number of times is related to the area of the Wilson loop. 
For simplicity we consider the case where the vortex surface intersects the plane in two nearby points separated by a distance $q$, one where it enters the plane~(marked in green) and the other where it exits~(marked in red). 
This visualization is shown in the second row.}
\end{figure}
To understand why the Wilson loop does not obey an intuitive expectation based on na\"{i}ve orientation counting. 
When the loop lies in a single plane, we analyze the role of center-vortex percolation.

In four-dimensional spacetime, center vortices are closed two-dimensional surfaces~\cite{Reinhardt:2001kf}. 
The Wilson loop lies in a two-dimensional plane. 
Generically, the intersection of two two-dimensional surfaces in four dimensions consists of isolated points. 
Because the vortex surface is closed, these intersections occur in pairs corresponding to entry and exit of the surface through the plane.
For an infinite, boundless plane, any closed vortex surface must intersect the plane an even number of times, so that the net center flux through the plane vanishes.

For a finite surface with boundaries~(corresponding to a Wilson loop), this topological constraint no longer forbids odd numbers of intersections, because the boundary allows vortex surfaces to intersect the minimal area an unpaired number of times. 
When the scale of the center vortex is small, only vortices lying within a distance comparable to their size from the loop boundary can contribute such odd intersections, leading to a boundary-dominated~(perimeter) contribution to the Wilson loop.
If vortex surfaces form only small closed loops, then only those lying near the boundary of the Wilson loop can produce such odd intersections, and the contribution to the Wilson loop is therefore dominated by the boundary, leading to a perimeter-type behavior.
If instead the vortex surfaces form very large connected structures that extend across the system, intersections are no longer confined to the boundary region but occur throughout the interior of the Wilson surface.
In this case, the number of odd intersections scales with the area of the surface, producing the area law for the Wilson loop.
This provides a simple geometric picture of how center vortices naturally lead to either perimeter behavior or area-law behavior of Wilson loops.
An example in three-dimensional spacetime~(in this case, the center vortices are closed loops) is shown in Fig.~\ref{fig:percolationpicture}.

For the purpose of constructing a simple qualitative model, we consider vortex configurations in which a vortex surface intersects the plane of the Wilson loop in a small number of localized regions. 
In the simplest case this produces a pair of nearby intersection points corresponding to entry and exit of the vortex through the plane. We model such configurations as pairs of points separated by a distance $q$
This physical representation is illustrated in the second row of Fig.~\ref{fig:percolationpicture}. 
It helps clarify why, when confined to the same plane, $\langle W_{s.o.}\rangle > \langle W_{o.o}\rangle$, indicating that the Wilson loop does not simply obey the na\"{i}ve orientation counting.
Examining cases where the Wilson loop is penetrated by up to two such vortex entries, six distinct scenarios arise. 
Denoting an entry into the plane as `+', an exit as `-', and no contribution as `0', these cases are: $(0,0)$, $(+,0)$, $(-,0)$, $(+,-)$, $(+,+)$, and $(-,-)$. 
The first three contribute equally to $\langle W_{s.o.}\rangle$ and $\langle W_{o.o}\rangle$, while the last three contribute oppositely to these averages.
If we model the center vortices as appearing in correlated pairs with endpoints separated by a distance $q$, then when one endpoint pierces the Wilson loop, the likelihood of an oppositely signed piercing occurring at another location roughly a distance $r\approx q$ away is slightly higher than that of a same-signed piercing.

\begin{figure}[htbp]
\includegraphics[width=0.49\hsize]{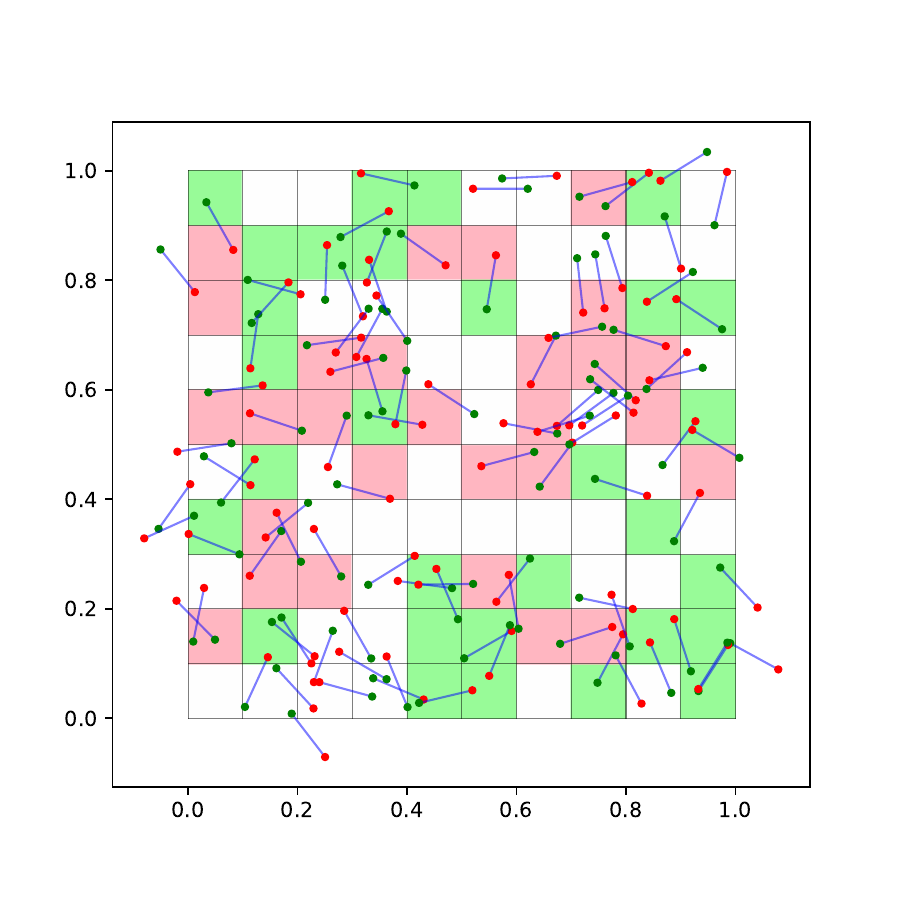}
\includegraphics[width=0.49\hsize]{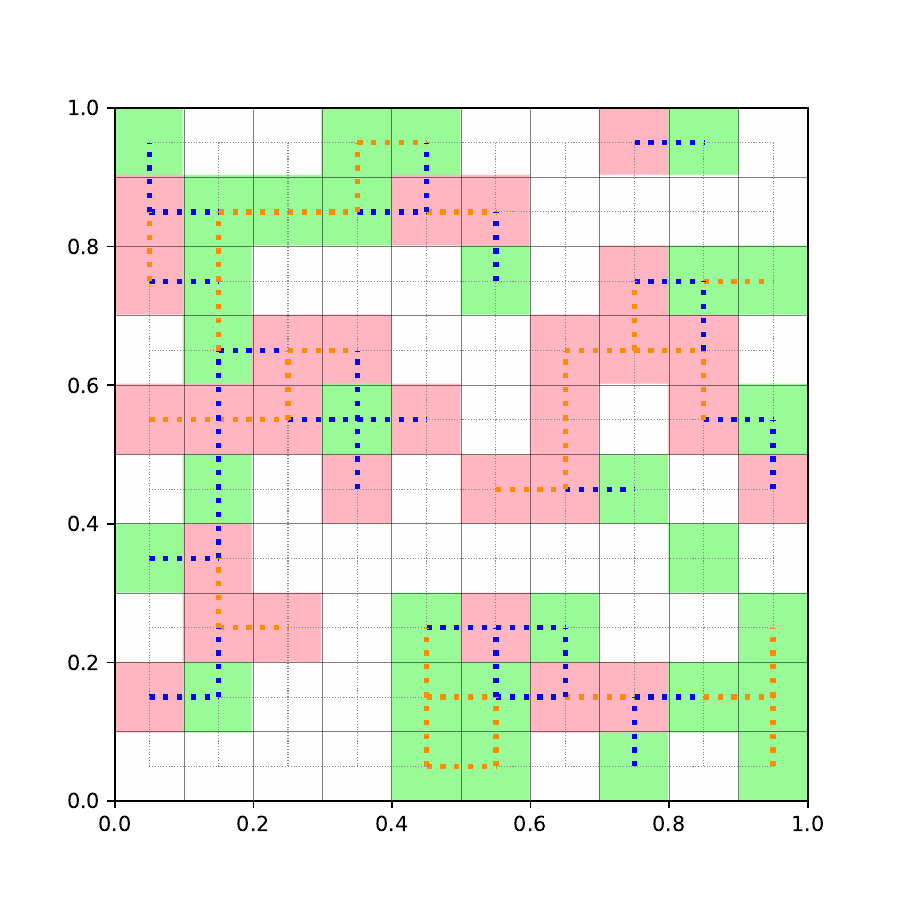}\\
\includegraphics[width=0.49\hsize]{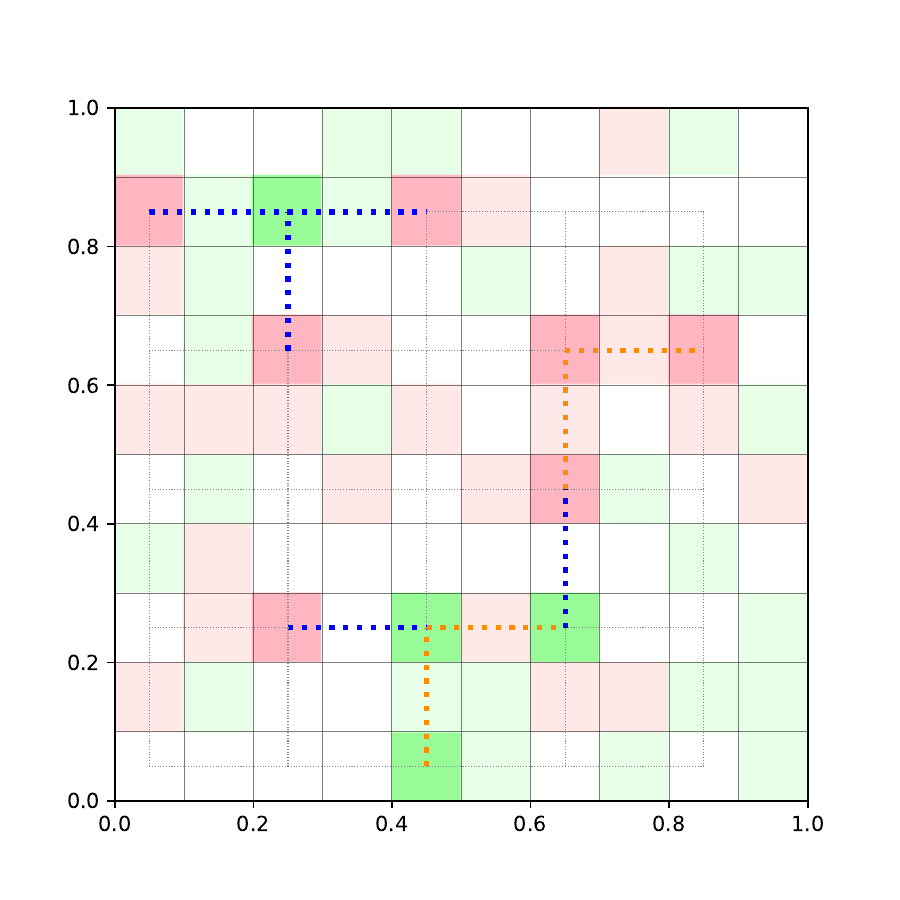}
\includegraphics[width=0.49\hsize]{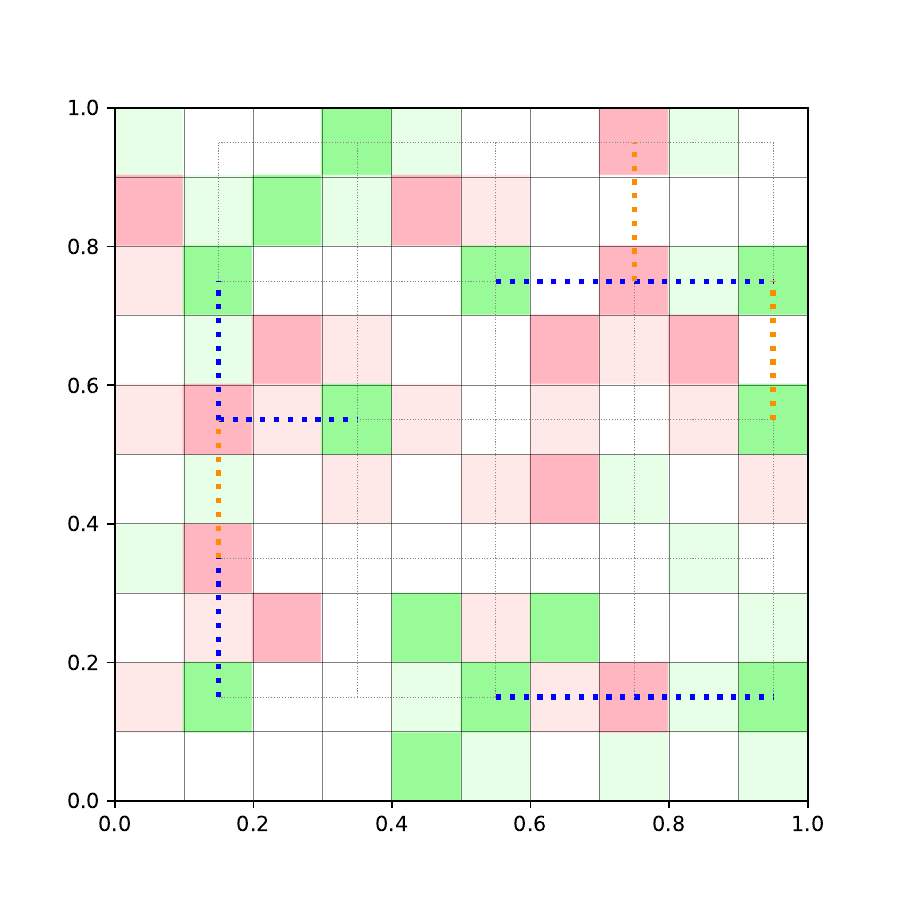}\\
\caption{\label{fig:percolationexample}
Assuming center vortices are randomly distributed line segments of length $q$ within a plane, with each of their two ends randomly assigned as either `+'~(marked in green) or `-'~(marked in red), as shown in the top-left panel. 
After generating such a random distribution, we divide the unit square into a grid of $k \times k$ small squares (here $k=10$ in the illustration) and compute the contribution to each small square as a Wilson loop. 
Small squares with phases `+', `-', and `0' are colored green, red, and white, respectively.
After determining the phase of each small-square Wilson loop, we calculate the phase relationship between two adjacent loops and record it as a link. 
As shown in the top-right panel, a blue link indicates that the two adjacent Wilson loops have opposite phases, a red link indicates they have the same phase, while all other cases are marked in gray.
Similarly, as illustrated in the second row, we compute the phase difference between two Wilson loops separated by one grid cell and again mark the corresponding links as either blue or red.}
\end{figure}

To verify this idea, we assume that the center vortices are randomly distributed line segments of a same length $q$ within a plane, with their two ends randomly assigned as `+' or `-'. 
As shown in the top-left panel of Fig.~\ref{fig:percolationexample}, we assume there are $m$ such segments with at least one endpoint lying inside a unit square. 
After randomly generating such a distribution, we divide the unit square into a grid of $k \times k$ small squares. 
For each small square, we compute the contribution (recorded as $0, +1, -1$) it receives as a Wilson loop.
After calculating the phases for all small-square Wilson loops, we analyze the phase relationship between two adjacent Wilson loops and record it as a link. 
As illustrated in the top-right panel of Fig.~\ref{fig:percolationexample}, a blue link indicates that the phases of two adjacent Wilson loops have opposite signs, while a red link indicates they have the same sign. 
Other cases contribute equally to $\langle W_{s.o.}\rangle$ and $\langle W_{o.o.}\rangle$ and are therefore not considered. 
We can then count the difference between the ratios of blue~(denoted as $R_{neg}^{(1)}$) and red links~(denoted as $R_{pos}^{(1)}$) to determine whether $\langle W_{s.o.}\rangle$ or $\langle W_{o.o.}\rangle$ is larger when two Wilson loops are adjacent.
Similarly, as shown in the second row of Fig.~\ref{fig:percolationexample}, we can calculate the phase difference between two Wilson loops separated by one grid cell. 
By counting the difference between blue and red links~(denoted as $R_{neg}^{(2)}$ and $R_{pos}^{(2)}$) at this separation, we can assess which of $\langle W_{s.o.}\rangle$ or $\langle W_{o.o.}\rangle$ is larger when the two Wilson loops are a small distance apart.

\begin{figure}[htbp]
\includegraphics[width=0.49\hsize]{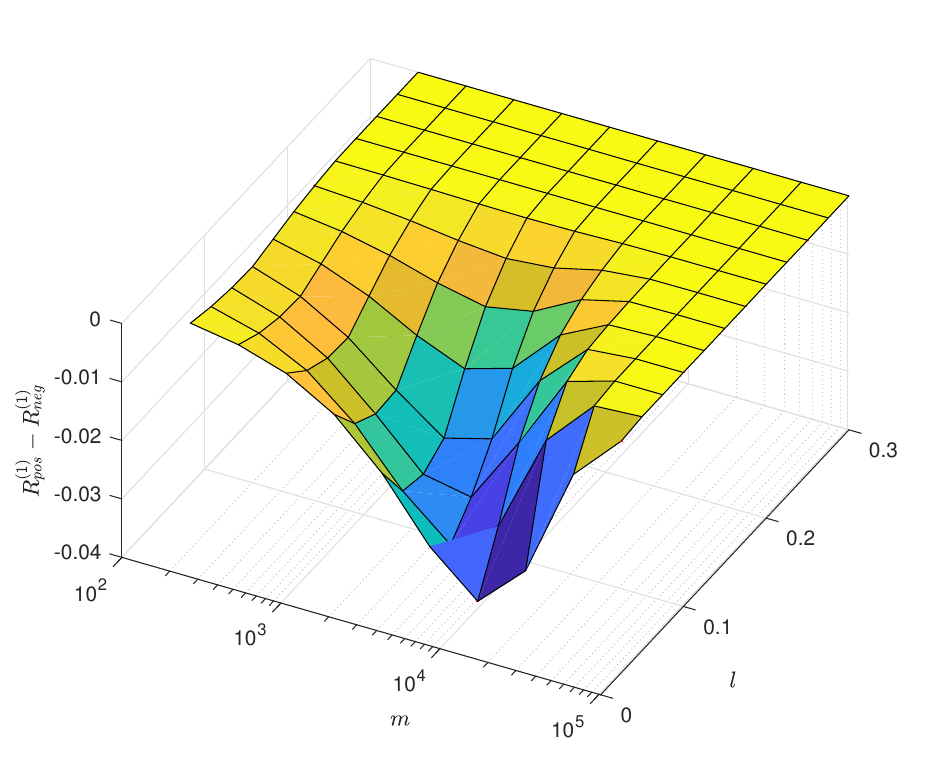}
\includegraphics[width=0.49\hsize]{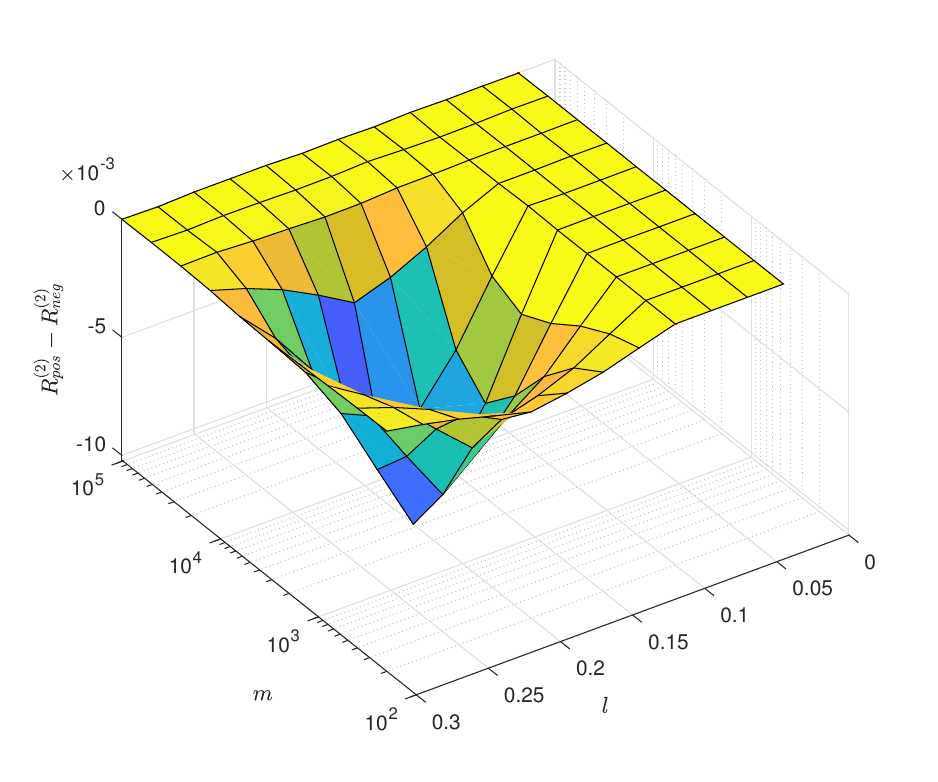}
\caption{\label{fig:percolation}
$R_{\text{pos}}^{(1)}-R_{\text{neg}}^{(1)}$~(left panel) and $R_{\text{pos}}^{(2)}-R_{\text{neg}}^{(2)}$~(right panel) as functions of $m$ and $r$.
In both cases there is a region that $R_{\text{pos}}^{(1,2)}-R_{\text{neg}}^{(1,2)}  < 0$, and $R_{\text{pos}}^{(1)}-R_{\text{neg}}^{(1)} < R_{\text{pos}}^{(2)}-R_{\text{neg}}^{(2)}$ which are consistent with the behavior of $\langle W_{s.o.}\rangle$ and $\langle W_{o.o.}\rangle$ when the Wilson loop is in a plane.}
\end{figure}
Since this na\"{i}ve physical picture is scale-invariant~(but depends on $l/q$), we are only concerned with relative scales. 
When $k=100$, the side length of each small square is $l=0.01$. 
We consider two cases: adjacent small-square Wilson loops (i.e., $r=0.01$) and Wilson loops separated by one grid cell (i.e., $r=0.02$). 
We also examine values of $q$ ranging from $0.0025$ to $0.03$, i.e. from $q=l/4$ to $q=3l$. 
For each case of given $m$ and $q$, we performed $10^6$ trials, and the resulting $R_{\text{pos}}^{(1)}-R_{\text{neg}}^{(1)}$ and $R_{\text{pos}}^{(2)}-R_{\text{neg}}^{(2)}$ are shown in Fig.~\ref{fig:percolation}. 
It can be observed that, within this parameter range, there indeed exists a region where the proportion of Wilson loops with opposite signs is significantly larger than that of Wilson loops with the same sign. 
In other words, when the areas of two Wilson loops are oriented in the same direction, they tend to receive cancelling contributions, while when the areas are oppositely oriented, they tend to receive reinforcing contributions. 
This leads to $\langle W_{o.o}\rangle < \langle W_{s.o.}\rangle $. 
We believe that the physical picture of center vortex penetration provides at least a qualitative explanation for the relationship between $\langle W_{o.o}\rangle$ and $\langle W_{s.o.}\rangle $.

\section{\label{sec5}Summary}

In this work we studied Wilson loops with a nontrivial orientation structure in lattice gauge theory in order to probe features of the center vortex picture of confinement. 
The observable considered here consists of a single Wilson loop containing two plaquettes with opposite orientations. 
Two geometric realizations of this construction are examined, referred to as the vertical and parallel configurations.
For the vertical configuration, the numerical results follow the qualitative expectations based on the vortex picture once the opposite orientations of the plaquettes are taken into account.

The parallel configuration, however, exhibits a qualitatively different behavior. 
The measured Wilson loop does not follow the simple area-law expectation, and this deviation cannot be explained solely by the opposite orientations of the plaquettes. 
This behavior therefore provides a nontrivial test of the vortex interpretation.

To understand this feature, we introduced a simple qualitative vortex model. 
The model suggests that correlations between vortex piercings associated with different parts of the loop can modify the naive area-law expectation and qualitatively account for the behavior observed in the parallel configuration. 
In this way the apparent anomaly can still be interpreted within the framework of the center vortex picture.

Although the present study is exploratory, the results indicate that Wilson loops with nontrivial orientation structures can serve as useful probes of vortex dynamics and correlations in the gauge vacuum.

\begin{acknowledgements}
This work was supported in part by the National Natural Science Foundation of China under Grants Nos. 12575106 and 12147214, the Basic Research Projects of Universities in Liaoning Province (Grant No.~LJKMZ20221431).
\end{acknowledgements}



\bibliography{vortexdirection}

\end{document}